\documentclass[12pt]{iopart}
\usepackage[english]{babel}
\usepackage{graphicx}
\usepackage{subfigure} 
\usepackage{iopams} 
\usepackage{float}
\usepackage{amssymb}

\usepackage{color}
\definecolor{darkgreen}{rgb}{0,0.6,0}
\definecolor{darkblue}{rgb}{0,0,0.6}
\definecolor{darkred}{rgb}{0.6,0,0}
\definecolor{darkpurple}{rgb}{0.5,0,0.5}
\usepackage[colorlinks=true,urlcolor=darkblue,citecolor=darkblue,linkcolor=darkred,hyperfootnotes=false]{hyperref}

\newcommand{\gap}{\textnormal{\scriptsize{gap}}}
\newcommand{\num}{\textnormal{\scriptsize{num}}}
\newcommand{\new}{\textnormal{\scriptsize{new}}}
\begin{document}

\title{Discreteness Effects in Population Dynamics}

\author{Esteban Guevara Hidalgo$^{1,2}$, Vivien Lecomte$^2$}

\address{$^1$Institut Jacques Monod, CNRS UMR 7592, Universit\'e Paris Diderot,
Paris Cit\'e Sorbonne, F-750205, Paris, France}
\address{$^2$Laboratoire Probabilit\'e  et Mod\`eles Al\'eatoires, CNRS UMR 7599, Universit\'e Paris Diderot, Paris Cit\'e Sorbonne, B\^atiment Sophie Germain,
Avenue de France, 75013 Paris, France}
 
\ead{esteban\_guevarah@hotmail.com} 
\vspace{10pt}

\begin{abstract}
We analyse numerically the effects of small population size in the initial transient regime of a simple example population dynamics. These effects play an important role for the numerical determination of large deviation functions of additive observables for stochastic processes. A method commonly used in order to determine such functions is the so-called cloning algorithm which in its non-constant population version essentially reduces to the determination of the growth rate of a population, averaged over many realizations of the dynamics. However, the averaging of populations is highly dependent not only on the number of realizations of the population dynamics, and on the initial population size but also on the cut-off time (or population) considered to stop their numerical evolution. This may result in an over-influence of discreteness effects at initial times, caused by small population size. We overcome these effects by introducing a (realization-dependent) time delay in the evolution of populations, additional to the discarding of the initial transient regime of the population growth where these discreteness effects are strong. We show that the improvement in the estimation of the large deviation function comes precisely from these two main contributions.

\vspace{2pc}
\noindent{\it Keywords}: Cloning Algorithm, Large Deviation Function, Population Dynamics, Birth-Death Process, Biased Dynamics, Numerical Approaches

\end{abstract}

%

%
%
%
%
%

\section{Introduction}

The occurrence of rare events can vastly contribute to the evolution of physical systems because of their potential dramatic effects.
Their understanding has gathered a strong interest and, focusing on stochastic dynamics, a large variety of numerical methods have been developed to study their properties.
They range from transition path sampling~\cite{Cochran_TPS,Hedges1,Speck} to ``go with the winner'' algorithms~\cite{Aldous,Grassberger} and discrete-time~\cite{GKP} or continuous-time~\cite{7} population dynamics (see~\cite{9} for a review), and they have been generalized to many contexts~\cite{delmoral,HGprl09,HGjstat09,lelievre,Vanden}.
In Physics, those are being increasingly used in the study of complex systems, for instance in the study of current fluctuations in models of transport~\cite{DerridaLebowitz,Derrida,MFT}, glasses~\cite{Hedges1}, protein folding~\cite{Weber1} and signalling networks~\cite{Vaikuntanathan,Weber2}.
Mathematically, the procedure amounts to determining a large deviation function (ldf) associated to the distribution of a given trajectory-dependent observable, which in turns can be reformulated in finding the ground state of a linear operator (see~\cite{hugoraphael} for a recent review of many aspects of this correspondence). In fact, this question is common to both statistical and quantum physics, and the very origin of population dynamics methods lies in the quantum Monte-Carlo algorithm~\cite{DMC}.

The idea of population dynamics is to translate the study of a class of rare trajectories (with respect to a determined global constraint) into the evolution of several copies of the original dynamics, with a local-in-time selection process rendering the occurrence of the rare trajectories typical in the evolved population.
The decay or growth of the population is in general exponential, at a rate which is directly related to the distribution of the class of rare trajectories in the original dynamics.
Two versions of such algorithms exist: the non-constant total population and the constant total population, for which a uniform pruning/cloning is applied on top of the cloning dynamics so as to avoid the exponential explosion or disappearance of the population.
While the later version is obviously more computer-friendly, the former version presents interesting features: First, it is directly related to the evolution of biological systems (stochastic jumps representing mutations, selection rules being interpreted as Darwinian pressure); Second, the uniform pruning/cloning of the population, although unbiased, induces correlations in the dynamics that one might want to avoid; Last, in some situations where the selection rates are very fluctuating, the constant-population algorithm cannot be used in practice because of finite-population effects (population being wiped out by a single clone), and one has to resort to the non-constant one.

In this article, we focus on the non-constant population algorithm, that we study numerically in a simple model where its implementation and its properties can be examined in great details.
In Section~\ref{The Cloning Algorithm and the Large Deviation Function}, we recall for completeness the relation between large deviations and the precise population dynamics.
In Section~\ref{sec:avepopldf} we describe issues related to the averaging of distinct runs,
that we quantify in Section~\ref{Parallel Behaviour in Log-Populations}.
In Section~\ref{sec:time_correct} we propose a new method to increase the efficiency of the population dynamics algorithm by applying a realization-dependent time delay, and
we present the results of its application in Section~\ref{sec:psi_timedelay}.
We characterize numerically the distribution of these time delays in Section~\ref{sec:time-delay-prop}.
Our conclusions and perspectives are gathered in Section~\ref{sec:discussion}.

\section{The Cloning Algorithm and Large Deviation Functions}
\label{The Cloning Algorithm and the Large Deviation Function}
A method commonly used in order to determine the large deviation function is the so-called ``cloning algorithm''~\cite{7,9,8}. This method has its origin in the study of continuous-time Markov chains and their dynamics. In this section we make a review of the theoretical background behind the algorithm, of how populations are generated and of how the ldf is evaluated.
 
\subsection{Continuous-time Markov Chains and the s-modified Dynamics}
Let $\{C \}$ be the set of possible configurations of a system which evolves continuously in time with jumps from $C$ to $C'$ occurring at transition rate $W(C \rightarrow C')$. The probability $P(C,t)$ to find the system at time $t$ in configuration $C$ evolves in time following the master equation
\begin{equation}\label{eq:1}
\partial_{t}P(C,t)={\sum\limits_{C' \neq C}^{ }} W(C'\rightarrow C)P(C',t) - r(C)P(C,t)
\end{equation}
where $r(C) = {\sum\limits_{C' \neq C}^{ }}  W(C \rightarrow C')$ is the escape rate from configuration $C$. 
If we define an additive observable $A$ over trajectories of the system (extensive in time, such as the number of configuration changes along the trajectory), which increases by an amount $\alpha(C,C')$ each time the system changes from $C$ to $C'$, the probability $P(C,t)$ can be detailed into $P(C,A,t)$. This probability is defined as the probability of finding the system at time $t$ in the configuration $C$ and with a value $A$ of the observable. In this case
\begin{equation}
\partial_{t}P(C,A,t)={\sum\limits_{C' \neq C}^{ }} W(C'\rightarrow C)P(C',A-\alpha(C',C),t) - r(C)P(C,A,t)
\end{equation}
We can bias the statistical weight of histories by introducing a parameter $s$ which fixes the average value of $A$, such that $s\neq 0$ favors its non-typical values ($s = 0$ characterizes the non-biased case) and for $ \hat{P} (C,s,t)={\sum\limits_{A}^{ }} e^{-sA}P(C,A,t)$,

\begin{equation}
\partial_{t} \hat{P}(C,s,t)={\sum\limits_{C' \neq C}^{ }} W_{s}(C'\rightarrow C)\hat{P}(C',s,t) - r(C)\hat{P}(C,s,t)
\end{equation}
where $W_{s}(C'\rightarrow C) = e^{-s \alpha(C',C)}W(C'\rightarrow C)$ are the ``$s$-modified'' rates. This new stochastic process is called ``$s$-modified dynamics'' and can be conveniently rewritten as 
\begin{equation} \label{eq:4}
\partial_{t} \hat{P}(C,s,t)={\sum\limits_{C' \neq C}^{ }} (\mathbb{W}_{s})_{CC'}\hat{P}(C',s,t) + \left[ r_{s}(C)- r(C)\right] \hat{P}(C,s,t)
\end{equation}
where 
\begin{equation} \label{eq:5}
(\mathbb{W}_{s})_{CC'} = W_{s}(C'\rightarrow C) - r_{s}(C)\delta_{CC'}
\end{equation}
and $r_{s}(C)=\sum\limits_{C'}^{ } W_{s}(C\rightarrow C')$. 

Equation (\ref{eq:4}) can be seen as the evolution equation of the (non-conserved) probability $\hat{P}(C,s,t)$ with rates $W_{s}(C'\rightarrow C)$, supplemented with a population dynamics where configurations are multiplied at a rate $\left[ r_{s}(C)- r(C)\right]$. The conjunction of these ``mutation'' and ``selection'' processes constitutes the cloning algorithm. The first term of the rhs conserves the probability while the second term represents the creation or destruction of clones of the system. The non-probability preserving ``selection'' process allows to render typical in the $s\neq 0$ biased dynamics an atypical class of histories of the original $s=0$ process. A detailed description of the corresponding cloning algorithm is given below.


\subsection{Populations and the Large Deviation Function} \label{Populations and the Large Deviation Function}
\subsubsection*{The Cloning Algorithm}
Consider $N_{0}$ clones of the system (\emph{i.e.}, $N_{0}$ copies of the system initially in the same configuration $C$). The dynamics is continuous in time and described by the times of configuration changes. We denote by  $ \mathbf t = \{ t^{(i)} \}_{i=1,...,N_{0}}$ the times at which each of the clones $ c = \{c_{i}\}_{i=1,...,N_{0}}$ will evolve. 

\begin{enumerate}
\item[1.] The first clone to evolve corresponds to the clone $c_{j}$ such that $t^{(j)} = \min \mathbf t$.
\item[2.] $c_{j}$ changes its configuration from $C$ to $C'$ with probability $W_{s}(C \rightarrow C')/r_{s}(C)$ 
\item[3.] $c_{j}$ is replaced by  $y = \lfloor Y(C) + \epsilon  \rfloor$ copies, where $Y(C) = e^{\Delta t(C) (r_{s}(C) - r(C))}$ is the cloning factor and $\epsilon$ is a random number uniformly distributed on $[0,1]$.
\item[4.] The next evolution of each copy $c_{j}$ will occur at $t^{(j)} + \Delta t(C')$ where $\Delta t(C')$ is chosen from a exponential law of parameter $r_{s}(C')$, and is drawn independently for every copy.
%
\item[5.]If $y = 0$, $c_{j}$ is erased. If $y>1$, we make $y-1$ copies of $c_{j}$.
 \end{enumerate}
 The repetition of this procedure will result (after an enough  time) in an exponential growth (or decay) of the number of clones. We restrict for simplicity our study to situations were the ldf is positive and the population thus increases in time. We can keep track of the different changes in the number of clones and of the times where these changes occur and we will denote by $N(s,t)$ the time-dependent population. Once we have generated $N(s,t)$, we can compute the ldf from the slope in time of the log-population $\hat{N}(s,t) = \log N(s,t)$, which constitutes an evaluation of the population growth rate. This can be done in different ways, for example by fitting $\hat{N}(s,t)$ by $y = mt+b$ and taking the ldf as $\Psi(s)=m$ or also by computing $\Psi(s)$ from
\begin{equation}\label{eq:6}
\Psi(s)=\frac{1}{T_{\max}-T_{\min}}\log\left(\frac{N_{\max}}{N_{\min}}\right)
\end{equation}
where $N_{\max}$ and $N_{\min} $ are the maximum and minimum values for $N(s,t)$ and $T_{\max}$ and $T_{\min}$ their respective times. We will refer later to this procedure as the 	``bulk'' slope estimator of the ldf.
 

Note that in some situations one can extend the previously described algorithm to keep population constant, by uniformly pruning or cloning the copies at each step so as to  effectively preserve the total population size without biasing its evolution. However, we are interested in situations where such approach cannot be applied in practice, for instance when the cloning rate is highly fluctuating.
 
\subsection{The Birth and Death Process}
\label{sec:bdp}

Throughout this article we focus our attention on a toy system where population discreteness can be studied simply: the birth and death process in one site. The system presents two states $0$ and $1$ and the transition rates read $W(0 \rightarrow 1) = c $ and $W(1 \rightarrow 0) = 1-c $ so that equation (\ref{eq:1}) for this process becomes
\begin{equation}\label{eq:7}
\partial_{t} \left(\begin{array}{c}
P(0,t)\\
P(1,t)
\end{array}\right) \left(\begin{array}{cc}
-c & 1-c\\
c & -1+c
\end{array}\right) \left(\begin{array}{c}
P(0,t)\\
P(1,t)
\end{array}\right)
\end{equation}
Additionally, for our purposes, we will consider as additive observable $A$ the activity $K$ for which $\alpha(C,C') = 1$: $K$ represents the total number of configuration changes up to final time $t$.
An advantage of considering this process for our analysis is that the large deviation function for the activity can be determined analytically. The large-time cumulant generating function $\Psi_{K}(s)=\lim_{t\rightarrow\infty}\frac{1}{t}\log \langle e{}^{-sK} \rangle$, also corresponds to the maximum eigenvalue of following matrix (see equation~(\ref{eq:5}))
\begin{equation}\label{eq:Ws}
\mathbb{W}_{s}=\left(\begin{array}{cc}
-c & (1-c)e^{-s}\\
ce^{-s} & -1+c
\end{array}\right)
\end{equation}
which it is found to be
\begin{equation} \label{eq:9}
\Psi_{K}(s)=-\frac{1}{2}+\frac{1}{2}\sqrt[]{1-4c(1-c)(1-e^{-2s})}
\end{equation}
Equation (\ref{eq:9}) will allow us to assess the quality of our numerical results.
The inverse of the difference between the eigenvalues of $\mathbb{W}_{s}$
\begin{equation} \label{eq:tgap}
t_{\gap} = \frac{1}{\sqrt{1-4c(1-c)(1-e^{-2s})}}    
\end{equation}
allows us to define the typical convergence time $t_{\gap}$ to the large time behaviour for equation~(\ref{eq:7}).

\section{Average Population and the LDF} 
\label{sec:avepopldf}
As we mentioned before, the cloning algorithm results (as time goes to infinity) in an exponential growth (for $s < 0$) or decay (for $s > 0$) of the number of clones. As we will see later, the ``discreteness effects'' in the evolution of our populations are strong at initial times. That is why the determination of the ldf using this algorithm is constrained not only to the parameters $(c,s)$, the initial number of clones $N_0$ and the number of realizations $R$ but also to the final time (or the maximum population) until which the process evolves in the numerical procedure.

In order to obtain an accurate estimation of~$\Psi(s)$, we should average several realizations of the procedure described in section~\ref{Populations and the Large Deviation Function}. To perform this average, we will define below a procedure that we have called ``merging'' which will allow us to determine in a systematic way the average population from which we can obtain an estimation of the ldf~$\Psi(s)$.
Noteworthy, this erroneously could be seen as obtaining~$\Psi(s)$ from the growth rate of the {average} (or equivalently the sum) of several runs of the population dynamics. 
This procedure would be incorrect since it amounts to performing a single run of the total population of the different runs, with a dynamics that would partition the total population into \emph{non-interacting} sub-populations, while, as described in section~\ref{Populations and the Large Deviation Function}, the population dynamics induces effective interactions among the whole set of copies inside the population.
In fact, the right way of performing this numerical estimate comes from computing~$\Psi(s)$ from the average growth rate of several runs of the population \emph{i.e.}, from taking the average $\langle \log N(s,t)\rangle$ of the slopes of several $\log N(s,t)$ instead of the slope of $\log \langle N(s,t) \rangle$.
The two results differ in general since $\langle \log N(s,t)\rangle \neq \log \langle N(s,t) \rangle$. One can expect that the two results become equivalent in the large $N_0$ limit as the distribution of growth rate should become sharply concentrated around its average value; however, they are different in the finite $N_0$ regime that we are interested in.

\subsection{Populations Merging}
\label{Populations Merging}
Let's consider $J$ populations  $\mathcal{N} = \{ N_{1}(s,t),N_{2}(s,t),...,N_{J}(s,t) \}$. The average population is defined as $\langle \mathcal{N} \rangle = \langle N_{j}(s,t) \rangle_{j = 1}^{J}$. In order to compute $\langle \mathcal{N} \rangle$, we introduce a procedure that we have called ``merging'' of populations which is described below. 

Given $N_{i}(s,t)$ and $N_{j}(s,t)$ the result of merging these two populations $\mathcal{M}(N_{i},N_{j})$ is another population $N_{ij} = N_{i} + N_{j}$ which represents the total number of clones for each time where a change in population for $N_{i}$ and $N_{j}$ has occurred. If $\langle N_{ij} \rangle$ is the average population for $N_{i}$ and $N_{j}$, the merged population and the average population are related through $\langle N_{ij} \rangle = \frac{N_{ij}}{2}$. If we add, for example, to our previous result another population $N_{k}$, the result $\mathcal{M}(N_{ij},N_{k})$ is related to the average by  $\mathcal{M}(N_{ij},N_{k})=N_{ij}+N_{k}=N_{i}+N_{j}+N_{k} = N_{ijk} = 3\langle N_{ijk} \rangle$. 

These ``merging'' procedure can be done for each of the populations in $\mathcal{N}$ so that
 \begin{equation} \label{eq:10}
 \mathcal{M}[\mathcal{N}]=\mathcal{M}(\mathcal{M}(\mathcal{M}(...(\mathcal{M}(\mathcal{M}(N_{1},N_{2}),N_{3}),N_{4}),...),N_{J-1}),N_{J})
 \end{equation}
is the result of systematically merging all the populations in $\mathcal{N}$. The average population $\langle \mathcal{N} \rangle$ can be recovered from $\mathcal{M}[\mathcal{N}]$ as $\langle \mathcal{N} \rangle = (1/J) \mathcal{M}[\mathcal{N}]$. Similarly, in the case of log-populations, $\langle \hat{N} \rangle = \langle \hat{N}_{j}(s,t) \rangle_{j = 1}^{J}$ (where $\hat{N}_{j}(s,t) = \log N_{j}(s,t)$) is obtained from merging all the populations in $\hat{N} = \{ \hat{N}_{1}(s,t),\hat{N}_{2}(s,t),...,\hat{N}_{J}(s,t) \} $. $\Psi(s)$ is then computed from the slope of $\langle \hat{N} \rangle$ with $\langle \hat{N} \rangle = (1/J) \mathcal{M}[\hat{N}]$.

\subsection{Discreteness Effects at Initial Times}
\label{Discreteness Effects at Initial Times}
Issues can emerge in the determination of $\Psi(s)$ which are not only related to the dependence of the method in $N_{0}$ (the initial number of clones) and $J$ (the number of populations). At initial times there is a wide distribution of times at which the first series of jumps occurs. This means that fluctuations at initial times induce that some populations remain in their initial states longer than others, producing an effective delay compared to other populations that evolve faster in their initial regime. From a practical point of view, this can induce that the numerical determination of $\Psi(s)$ becomes a slow and inefficient task. One way of dealing with this issue comes from restricting the evolution of $\mathcal{N}$ up to a maximum time $T_{\max}$ or a maximum population $N_{\max}$. However, this implies that if $T_{\max}$ or $N_{\max}$ are not long enough, the determination of $\Psi(s)$ will be strongly affected by the behaviour of $\mathcal{N}$ at initial times. We now discuss two issues that are encountered in the numerical evaluation of the ldf: (i) the influence of how the dynamics is halted; and (ii) the role of initial population in the initial regime. 
 
\begin{figure}[h!]
        \centering
        \subfigure[$\langle \hat{N} \rangle_{[0,\min \mathcal{T_{F}}]}$] {\includegraphics [scale=0.51] {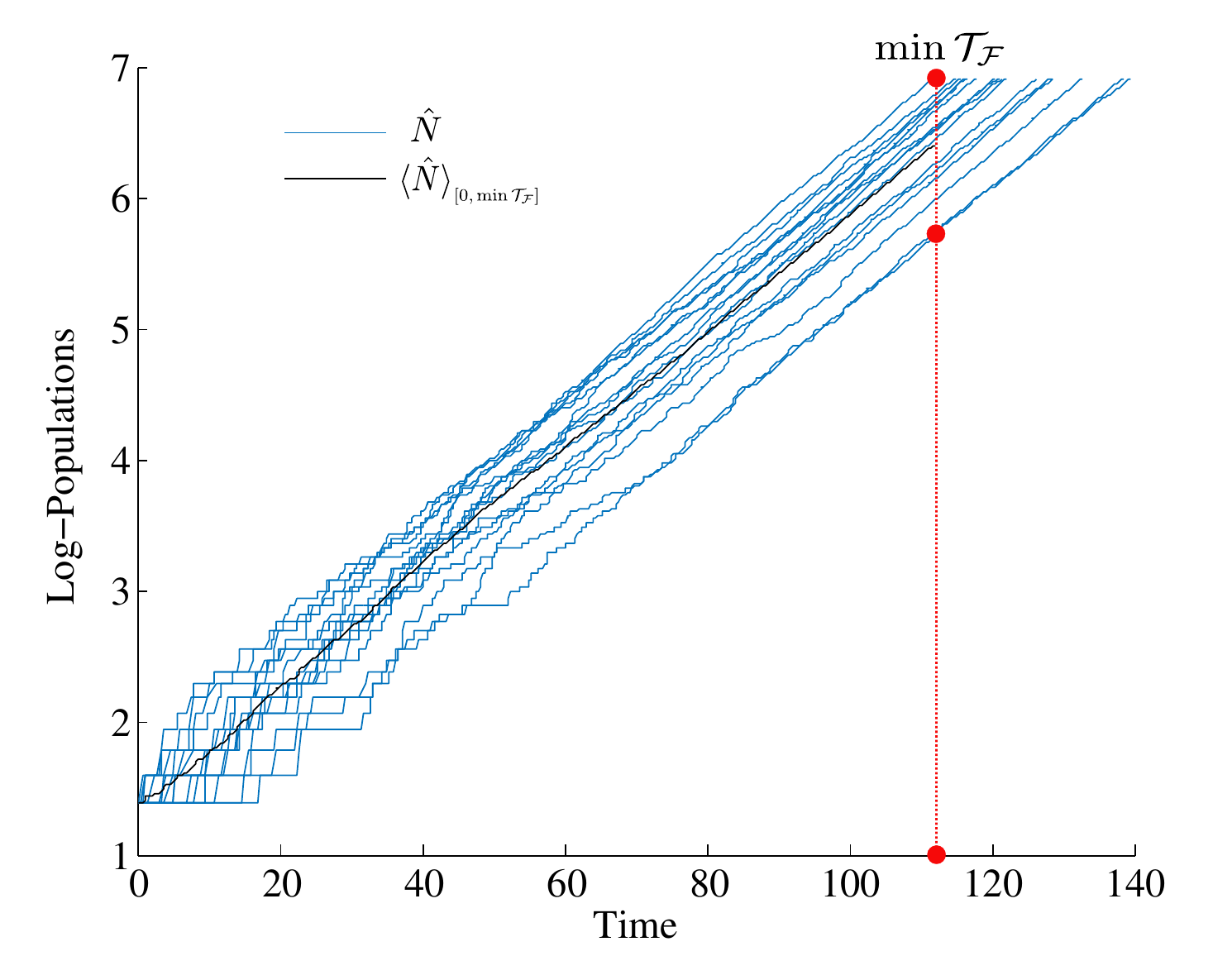}}
		\subfigure[$\langle \hat{N} \rangle_{[\max \mathcal{T_{C}},\min \mathcal{T_{F}}]}$]{\includegraphics [scale=0.51] {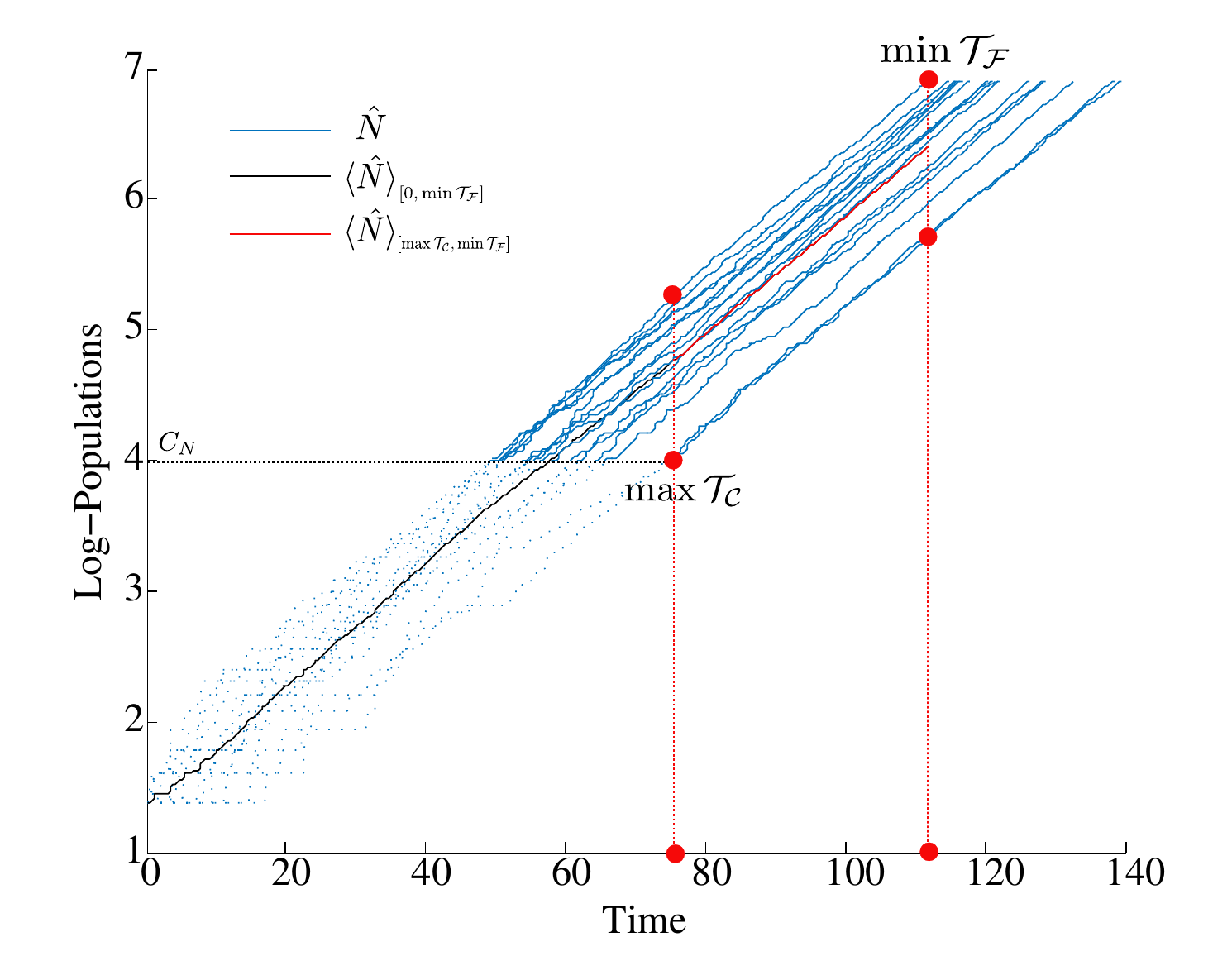}}\\ 
\centering        	
\caption{Log-populations as function of time (blue). Their evolution has been restricted up to a maximum (log) population value. (a) The average log-population $\langle \hat{N} \rangle$ (black) is made in the interval $[0,\min \mathcal{T_{F}}]$ where all the populations are defined. (b) After a cut in populations $C_{N}$ (in order to eliminate the initial discreteness effects), the average log-population (red) that represents the new $\hat{N}$ is defined only in the interval $[\max \mathcal{T_{C}},\min \mathcal{T_{F}}]$.}
\label{fig:merge}
\end{figure}

Let's call $\mathcal{T_{F}} = \{ t_{1}^{\mathcal{F}},...,t_{J}^{\mathcal{F}} \}$ the set of final times of $\mathcal{N}$,  with $t_{j}^{\mathcal{F}} \leq T_{\max}$, $\forall j\in\{1,...,J\}$. Note that $t_{j}^{\mathcal{F}}$ depends on $j$ whenever the simulation is stopped at $N_{\max}$ (as in figure~\ref{fig:merge}) or at $T_{\max}$. This due to the fact the algorithm is continuous in time and the last $\Delta t(C)$ does not exactly lead to $T_{\max}$. We say that the average population $\langle \mathcal{N} \rangle$ represents $\mathcal{N}$ only if the average is made in the interval $[ 0,\min \mathcal{T_{F}}]$ where all the populations are defined. In other words, the average population in this interval takes into consideration all the populations while for times $t \geq \min\mathcal{T_{F}}$ some populations have stopped evolving. This phenomenon is especially evident when considering a maximum population limit $N_{\max}$ for the evolution of the populations (Figure~\ref{fig:merge}(a)). As a consequence, $\langle \mathcal{N} \rangle$ depends on the distribution of final times of $\mathcal{N}$ which are not necessarily equal to $T_{\max}$. 
 
An alternative that can be considered in order to overcome the influence of initial discreteness effects in the determination of $\Psi(s)$ is to get rid of the initial transient regime where  these effects are present. In other words, to cut the initial time regime of our populations. Let's call $C_{N} \geq \log N_{0}$ the initial cut in log-populations and equivalently $C_{t} \geq 0$ the initial cut in times. $\mathcal{T_{C}} = \{ t_{1}^{\mathcal{C}},...,t_{J}^{\mathcal{C}} \}$ is the distribution of times at $C_{t,N}$. In that case, similarly as we analysed before, the average population $\langle \mathcal{N} \rangle$ represents $\mathcal{N}$ only if the average is made in the interval $[\max \mathcal{T_{C}}, \min \mathcal{T_{F}}]$ which can be in fact very small and could result in a bad approximation of $\Psi(s)$ (Figure~\ref{fig:merge}(b)).

\newpage
As we will see in the next section, log-populations after a long enough time become parallel \emph{i.e.}, once the populations  have overcome the discreteness effects regime, the distance between them is constant. We will use this fact in order to propose a method which allows us to overcome the problems we have described in this section. Throughout this article, we consider for our simulations $c = 0.3$, $N_{0} = 2^{2}$, $N_{\max} = 10^{3}$, $J = 2^{8}$ and $s \in \left[-0.3,0 \right] $.

\section{Parallel Behaviour in Log-Populations}
\label{Parallel Behaviour in Log-Populations}
\subsection{Distance between Populations}
Given $N_{i}(s,t)$ and $N_{j}(s,t)$, we define the distance between these populations at $N^{*}$ (with $N^{*} \in N_{i}$ and $N^{*} \in N_{j}$), as
\begin{equation} \label{eq:11}
D(N_{i},N_{j})(N^{*}) = \left \vert  \left( t_{j}(N^{*}) + \frac{\Delta t_{j}(N^{*})}{2} \right) - \left( t_{i}(N^{*}) + \frac{\Delta t_{i}(N^{*})}{2} \right) \right \vert
\end{equation}
where $\Delta t_{k}(N^{*})$ is the time interval $N_{k}(s,t)$ spent at $N^{*}$ and $t_{k}(N^{*})$ is the time where $N_{k}(s,t)$ changes to $N^{*}$. Evidently there are cases where $N^{*} \notin N_{i}$ but $N^{*} \in N_{j}$, $N^{*} \in N_{i}$ but $N^{*} \notin N_{j}$ and $N^{*} \notin N_{i}$ and $N^{*} \notin N_{j}$, however $D(N_{i},N_{j})(N^{*})$ for these cases can also be computed. The last analysis (and definitions) is also valid for log-populations. $D(N_{i},N_{j})(N^{*})$ and $D(\hat{N}_{i},\hat{N}_{j})(N^{*})$ enjoy interesting properties that we discuss below.

\begin{figure}[h!]
        \centering
        \subfigure[Log-Populations: $\hat{N}_{i}(s,t),\hat{N}_{j}(s,t)$]{\includegraphics [scale=0.51] {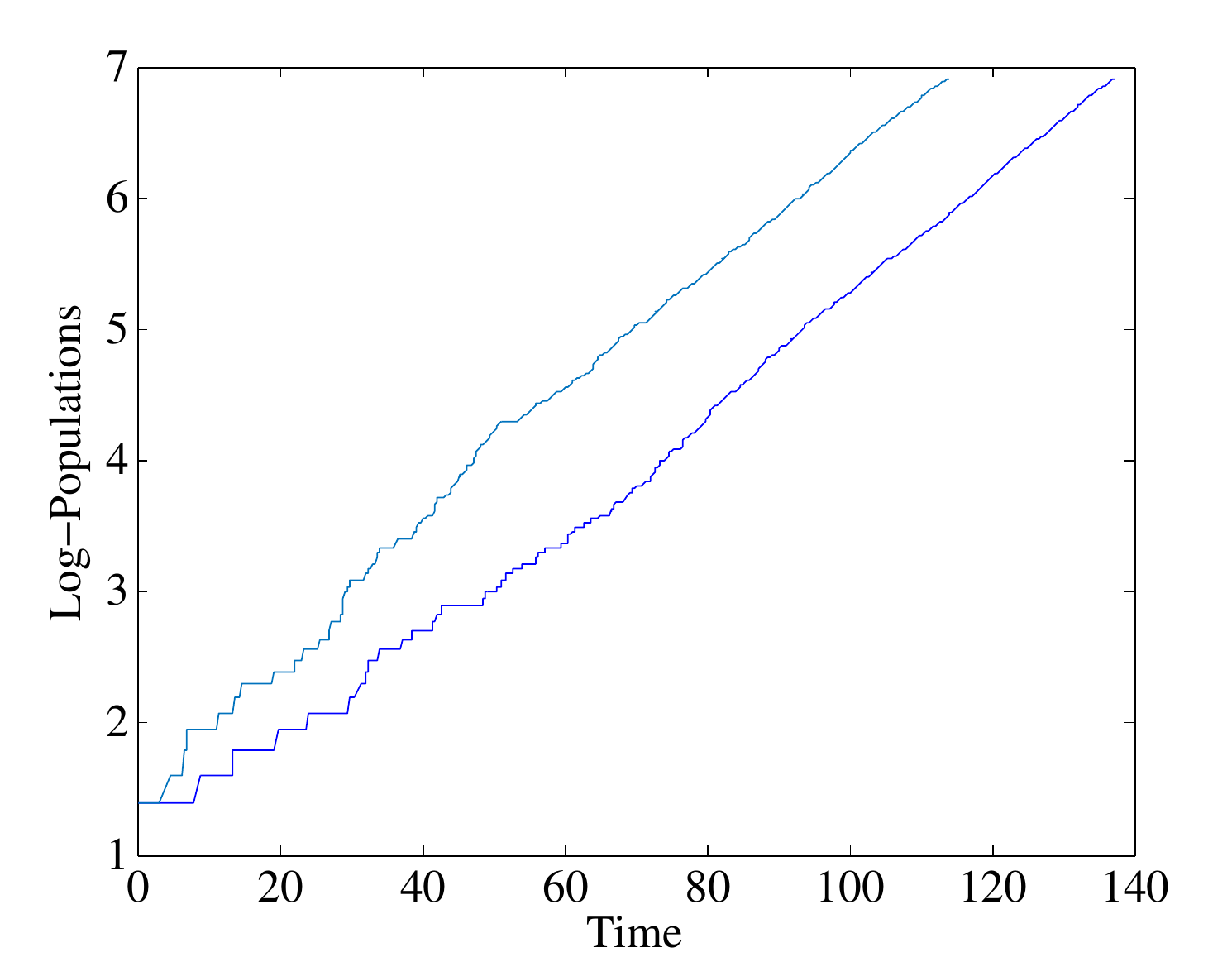}}
		\subfigure[$D(\hat{N}_{i},\hat{N}_{j})$]{\includegraphics [scale=0.51] {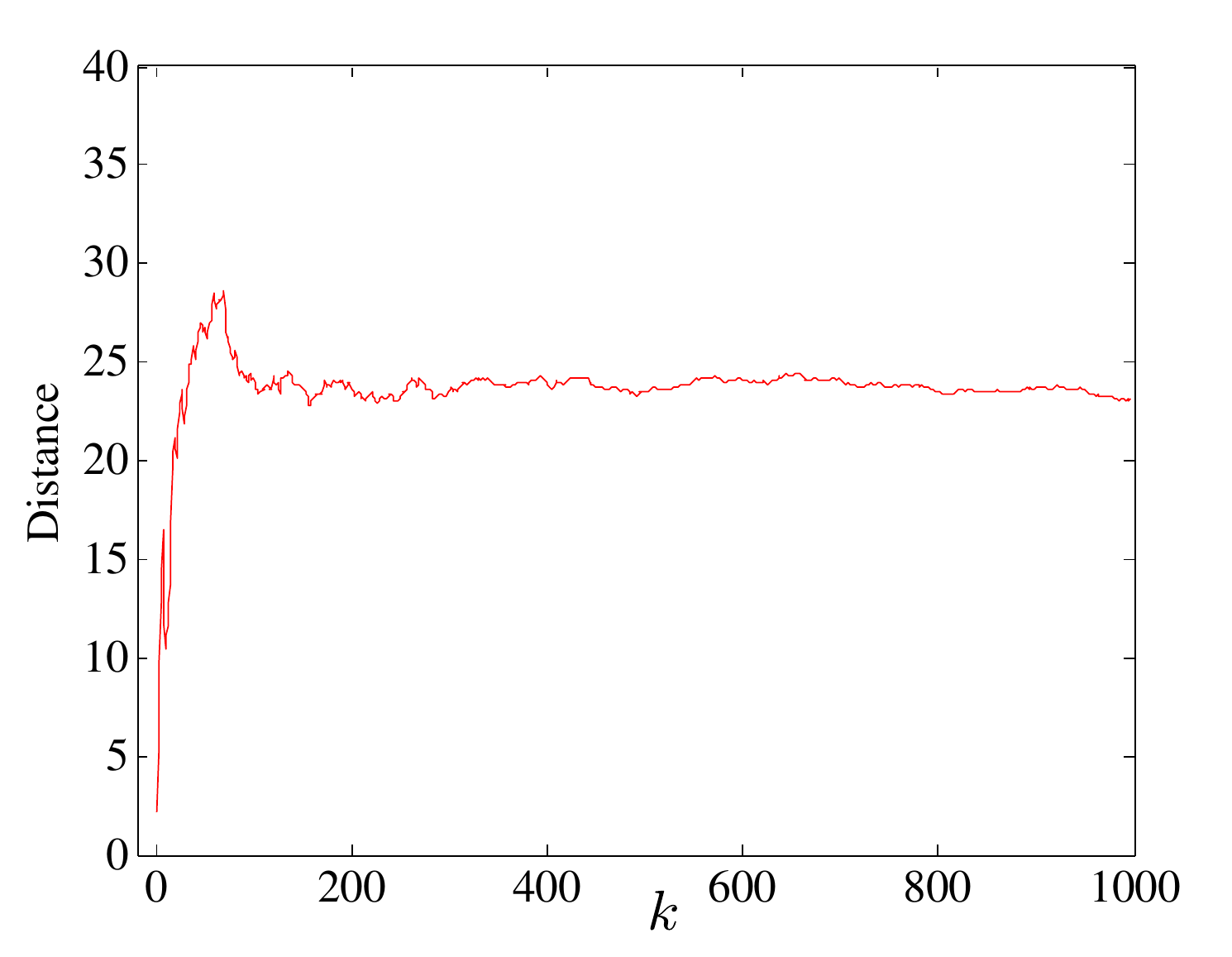}}\\ 
\centering        	
\caption{Evolution of two log-populations $\hat{N}_{i},\hat{N}_{j}$ as function of time and the distance $D(\hat{N}_{i},\hat{N}_{j})$ between them as defined in equation (\ref{eq:11}). (a) Log-populations after a long enough time become parallel. (b) Once the populations have overcome the initial discrete population regime, the distance between them becomes constant. $(s=-0.1)$. }
\label{fig:distance1}
\end{figure}

\subsection{Properties of $D(\hat{N}_{i},\hat{N}_{j})$}
In figure~\ref{fig:distance1}, we show two log-populations and the distance between them. As can be seen in figure~\ref{fig:distance1}(a) log-populations after a long enough time become parallel \emph{i.e.}, once the populations have overcome the discreteness effects, the distance between them becomes constant as can be seen in figure~\ref{fig:distance1}(b). The region where the distance between populations is constant characterizes the exponential regime of the populations growth, \emph{i.e.}, the region where the discreteness effects are not strong anymore.

\begin{figure}[h!]
        \centering
        \subfigure[$\langle D(\hat{N}_{F},\hat{N}_{j}) \rangle_{j}$] {\includegraphics [scale=0.51] {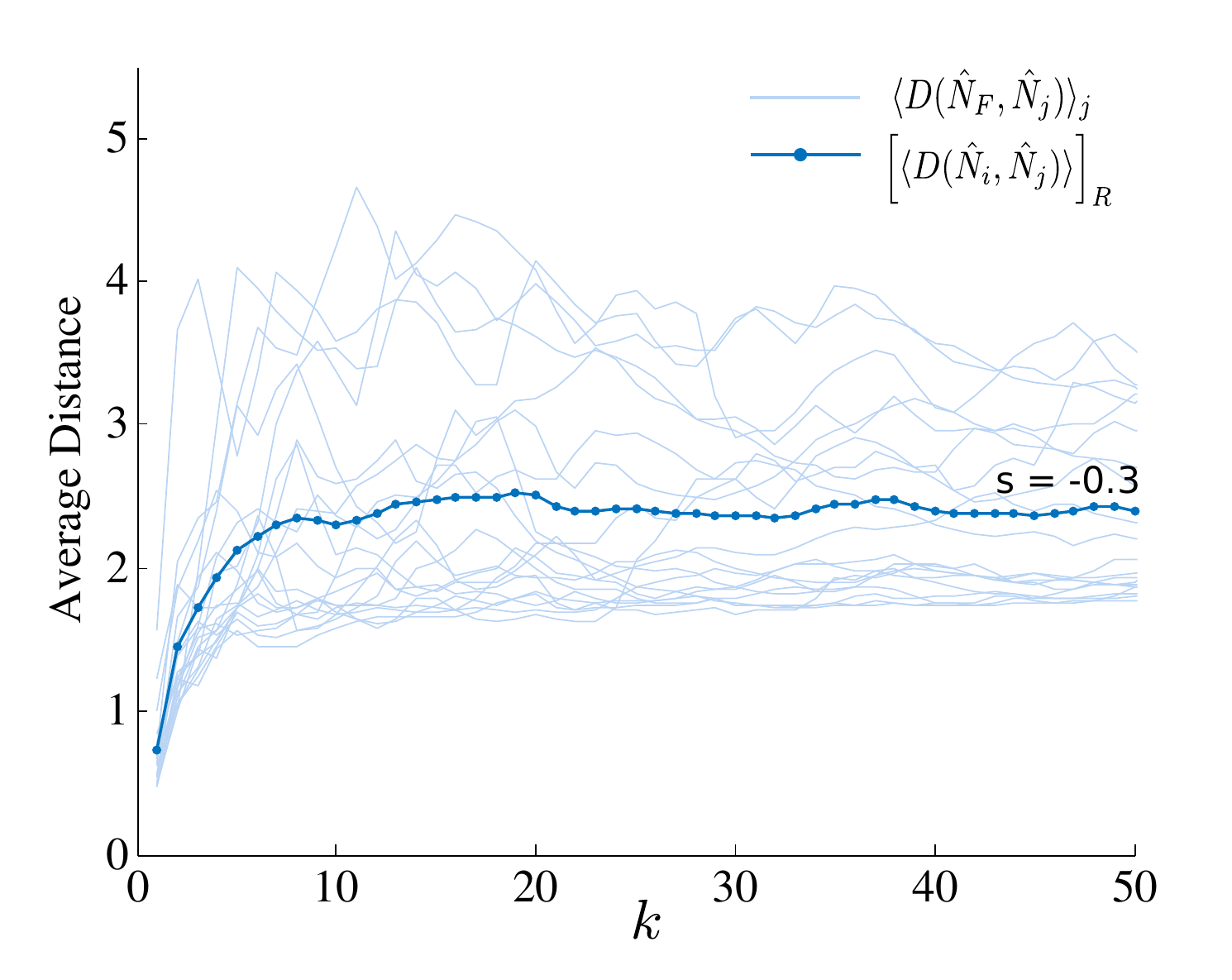}}
		\subfigure[Average $\langle D(\hat{N}_{F},\hat{N}_{j}) \rangle_{j}$]{\includegraphics [scale=0.51] {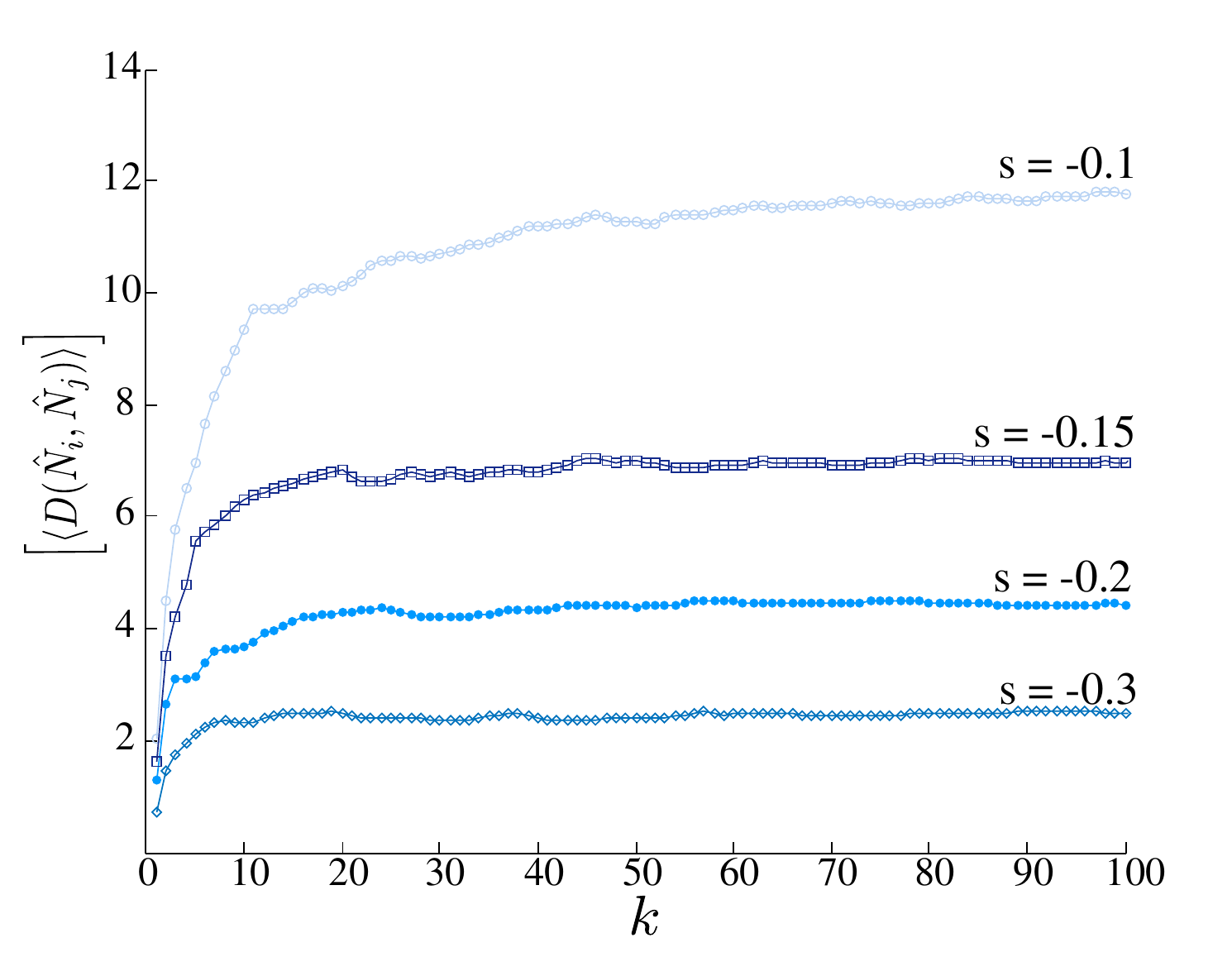}}\\ 
\centering        	
\caption{(a) Average distance of the populations in $\mathcal{N}$ with respect to a reference one for $R = 20$ realizations (light blue) and their average (dark blue). (b) Average distance between populations for several values of $s$. How fast a population ``exits" from the discreteness effects regime depends on $s$: $s$ closer to zero corresponds to a slower population growth and hence a longer discreteness regime.} 
\label{fig:distance2}
\end{figure}

If we consider some population $\hat{N}_{F} \in \hat{N}$ as reference, using the definitions above, it is possible to determine the distance $D(\hat{N}_{F},\hat{N}_{j})$ between $\hat{N}_{F}$ and the rest of populations in $\hat{N} = \{ \hat{N}_{1},\hat{N}_{2},...,\hat{N}_{J} \}$. In figure~\ref{fig:distance2}(a) we show their average $\langle D(\hat{N}_{F},\hat{N}_{j}) \rangle_{j}$ in light blue and its average over $R = 20$ realizations $\left[\langle D(\hat{N}_{i},\hat{N}_{j}) \rangle \right]_{R}$ in dark blue. As we mentioned in section~\ref{The Cloning Algorithm and the Large Deviation Function}, the parameter $s$ characterizes atypical behaviours of the unbiased dynamics, and this induces a dependence in $s$ of the population growth.  A population with a large value of $s$ corresponds to a large deviation of $K$. Also, as it is clearly illustrated in figure~\ref{fig:distance2}(b), the time of entrance of the system into a regime free of discreteness effects depends on $s$. 

\section{Time Correction in the Evolution of Populations}
\label{sec:time_correct}
Based on the results we just illustrated, we propose a method in order to improve the approximation of $\Psi(s)$ and reduce the influence of the initial discrete population size regime we described in subsection~\ref{Discreteness Effects at Initial Times}. We aim at giving more weight to the exponential regime in the determination of $\Psi(s)$. As we will see below, this can be done through a delay in the evolution of populations. 

\subsection{Time Delay Correction}
\label{Time Delay Correction}
\begin{figure}[h!]
        \centering
        \subfigure[Log-Populations] {\includegraphics [scale=0.51] {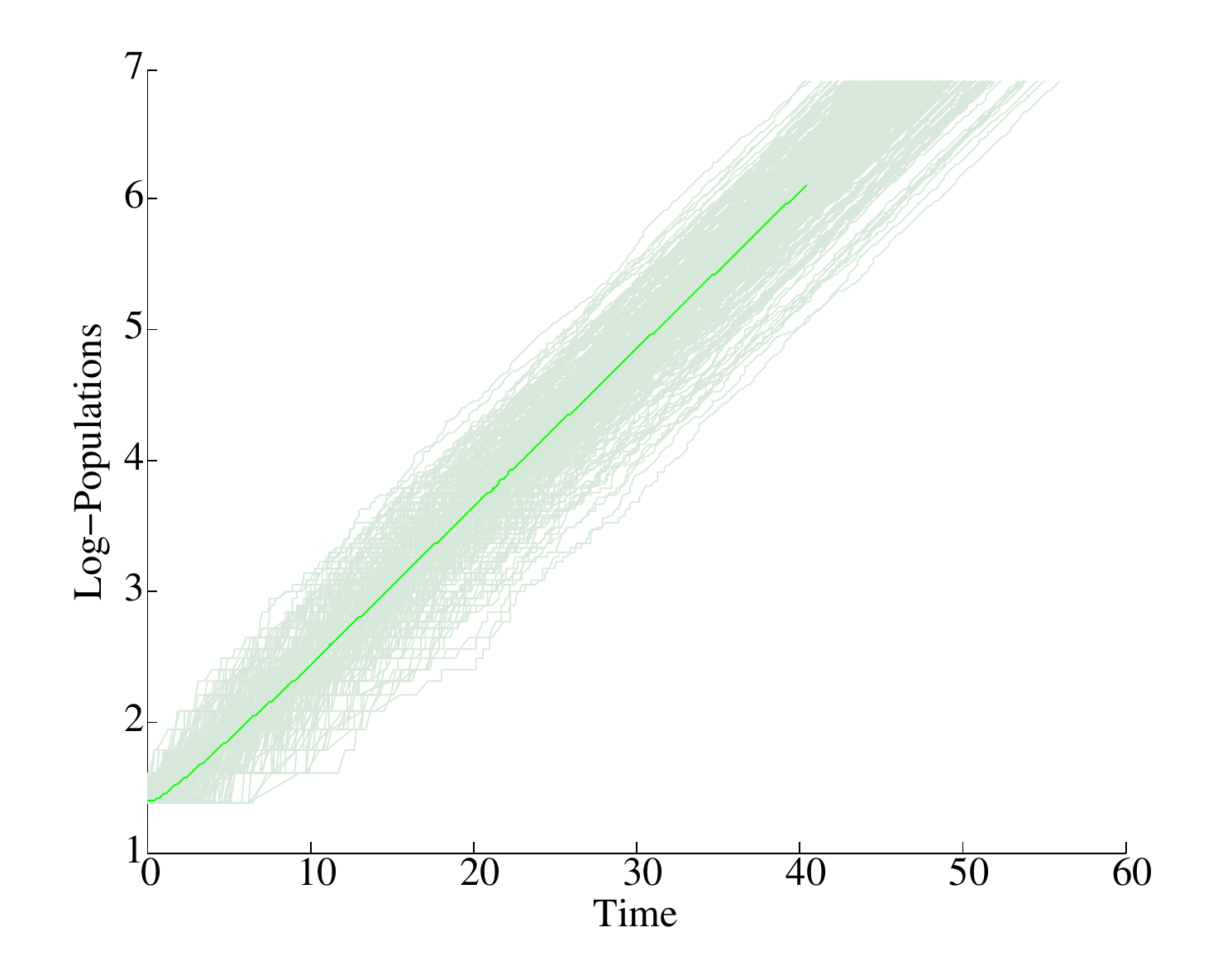}}
		\subfigure[Delayed Log-Populations]{\includegraphics [scale=0.51] {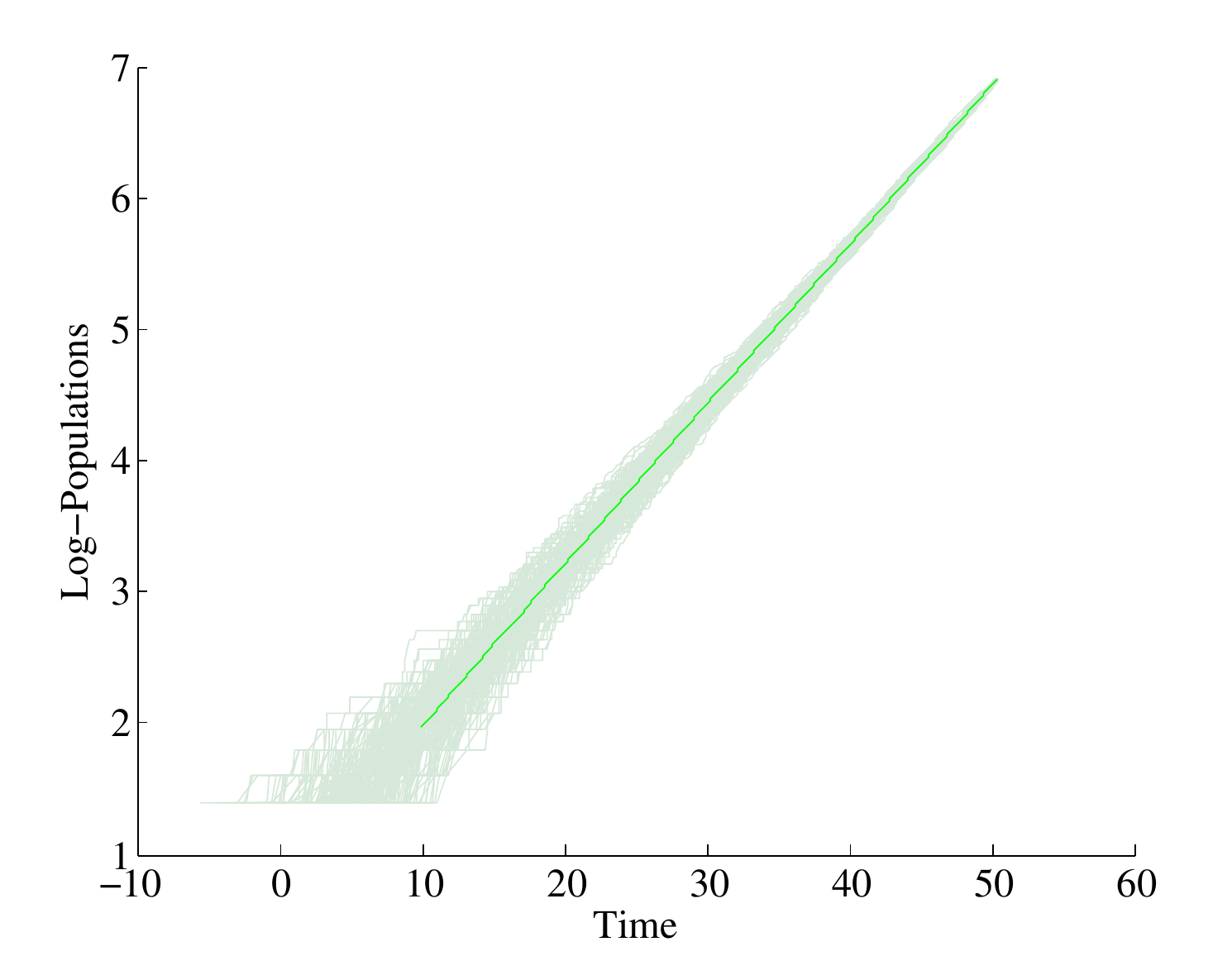}}\\ 
\centering        	
\caption{(a) Log-populations, (b) time-delayed log-populations, and their average (dark green). The fluctuations at initial times produce a gap in the evolution of individual populations inducing a relative shift that lasts forever. This is compensated by delaying the populations in time, as explained in section $4$. $(s=-0.25)$.}
\label{fig:delayedPOP}
\end{figure}

Consider $J$ populations $\mathcal{N}$, their respective log-populations $\hat{N}$ and their distribution of final times $\mathcal{T_{F}} = \{t_{1}^{\mathcal{F}},...,t_{J}^{\mathcal{F}} \}$. We define as ``delay" $\Delta\tau_{j}$ of $\hat{N}_{j}$ (with respect to a fixed reference population $\hat{N}_{F} \in \hat{N}$) the time interval

\begin{equation} \label{eq:12}
\Delta\tau_{j}=t_{F}^{\mathcal{F}} - t_{j}^{\mathcal{F}}
\end{equation}
such that, if $\Delta\tau_{j}<0$, $\hat{N}_{j}$ is ahead with respect to $\hat{N}_{F}$, and if $\Delta\tau_{j}>0$, $\hat{N}_{j}$ is delayed with respect to $\hat{N}_{F}$. This lag can be compensated by performing on $\hat{N}_{j}$ the time translation

\begin{equation} \label{eq:13}
\hat{N}_{j}^{\new} = \hat{N}_{j}(s,t+ \Delta\tau_{j})
\end{equation}
which produces that $\hat{N}_{j}^{\new}$ and $\hat{N}_{F}$ share not only the final population $N_{\max}$ but also the same final time $t_{F}^{\mathcal{F}}$. Moreover, considering also the fact that log-populations are parallel at large times, this procedure produces that $\hat{N}_{j}^{\new}$ and $\hat{N}_{F}$ overlap in the region that we have called free of discreteness effects. The result of performing this transformation to all the populations in $\hat{N}$ is shown in figure~\ref{fig:delayedPOP}.

Figure~\ref{fig:delayedPOP} also illustrates many of the points we have discussed up to now. One of them is related to the ``wide'' distribution of final times, \emph{i.e.}, $\min \mathcal{T_{F}}$ and $\max \mathcal{T_{F}}$ can be very distant one from each other. This along with the fact that the average population depends on $\min \mathcal{T_{F}}$ makes that the determination of $\Psi(s)$ omits a considerable region where the populations have already entered the exponential regime. This implies precisely that more weight is given to the initial discreteness effects than to the exponential regime. These effects are in fact present up to relatively long times which means that if we  would like to get rid of the region were discreteness effects are strong by cutting the populations, the determination of $\Psi(s)$ would be restricted to the interval $[\max \mathcal{T_{C}}, \min \mathcal{T_{F}}]$. 
By applying precisely this time delay correction to $\hat{N}$ we solve these two problems. First, we give more importance precisely to the region where the population growth is exponential. Second, we omit naturally the very first initial times of the evolution of our populations. 

As we mentioned in section~\ref{sec:bdp}, the inverse of the difference between the eigenvalues of $\mathbb{W}_{s}$ (equation~(\ref{eq:Ws})) $t_{\gap}$ allows us to define the typical convergence time to the large time behaviour for equation~(\ref{eq:7}). A crucial remark is that, as observed numerically, the duration before the population enters into the exponential regime is in fact larger than the time scale given by the gap: for instance, for the parameters used to obtain figure~\ref{fig:delayedPOP}, from equation~(\ref{eq:tgap}) one has $t_{\gap} \approx 0.804$. The understanding of the duration of this discreteness effects regime would require a full analysis of the finite-population dynamics and its associated discreteness effects, which are not fully understood. We propose in this section a numerical procedure to reduce its influence.

\subsection{Log-Population Variance}
\label{Log-Population Variance}

\begin{figure}[h!]
	 \centering
        \subfigure[Log-Population Variance] {\includegraphics [scale=0.51] {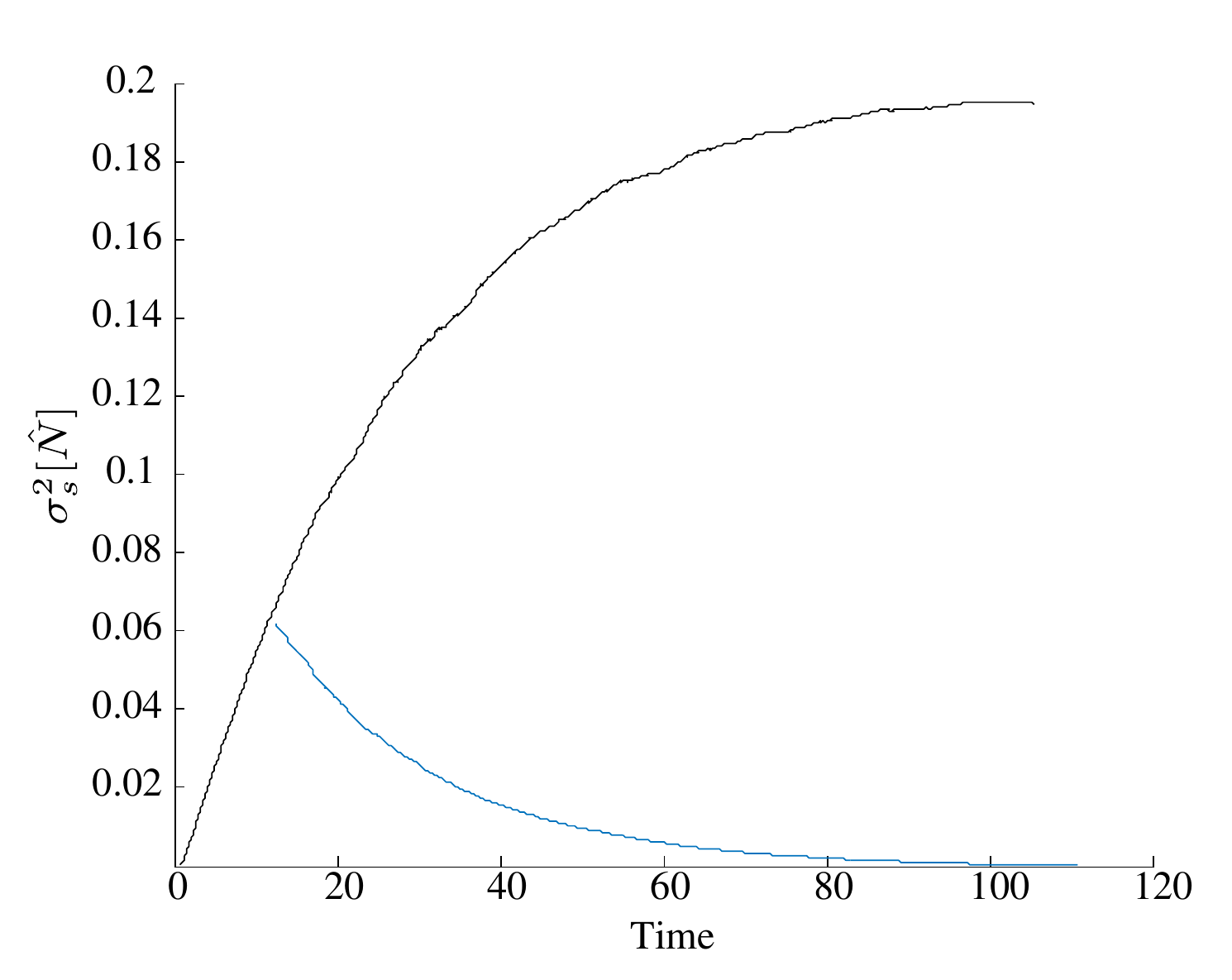}}
		\subfigure[ s - Dependent Decrease Rate]{\includegraphics [scale=0.51] {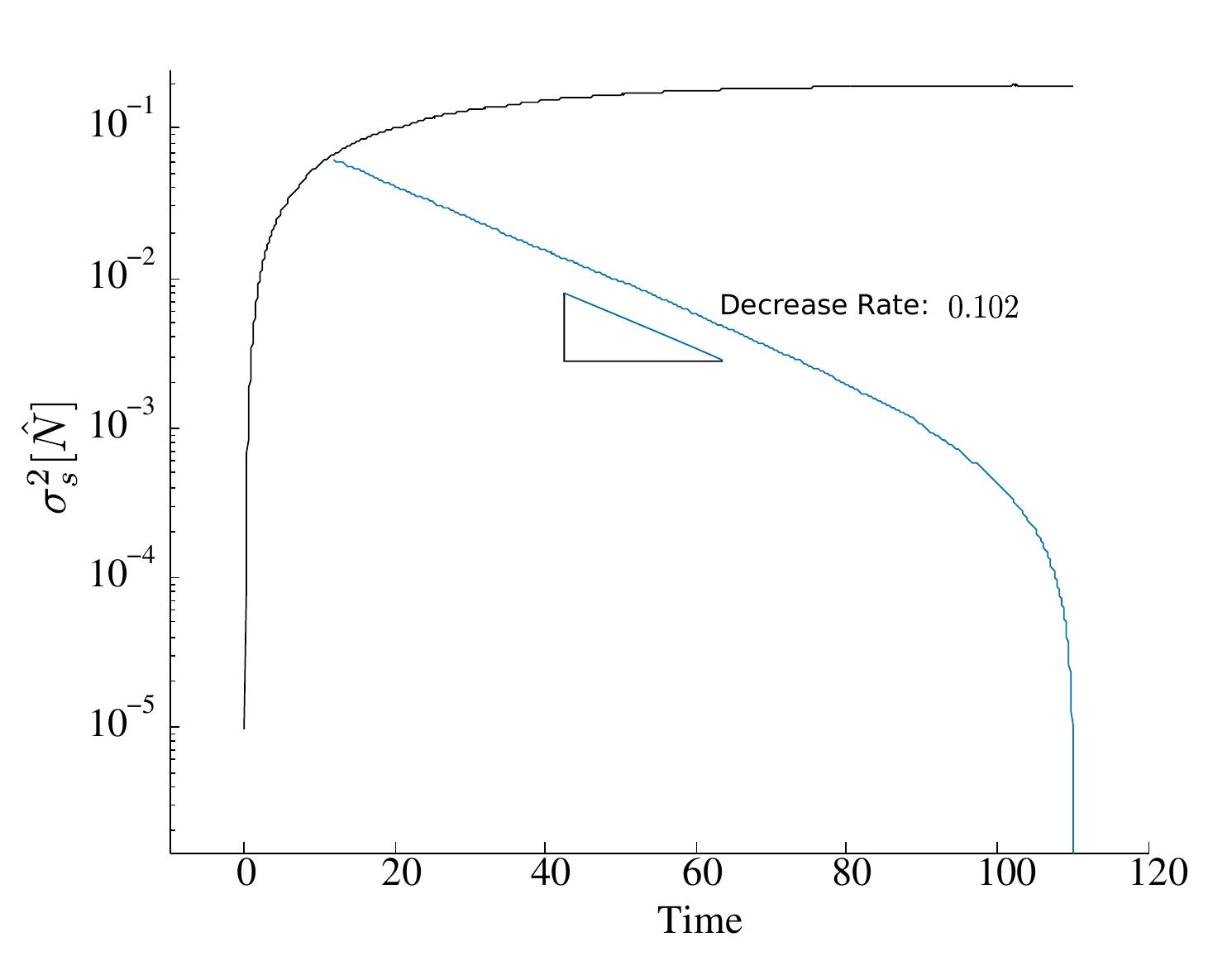}}\\ 
	\centering           	
\caption{(a) Variance of the log-populations (black) and the delayed log-populations (blue) as a function of time. The variance of log-populations increases (or decreases, after the time transformation) as function of time. (b) Log-population variance in semi-log scale. $(s = -0.1)$.}
\label{fig:varlogPop}
\end{figure}

As can be seen from figure~\ref{fig:delayedPOP}, and as it is proved in figure~\ref{fig:varlogPop}, the variance of log-populations (black) increases as a function of the time, faster during the transient regime, and slower during the exponential growth regime until the variance becomes constant. After the time-delay correction, the variance of the delayed log-population (blue) decreases to zero as a function of time. The $s$-dependent decrease rate is shown in figure~\ref{fig:varlogPop}(b).

\section{$\Psi(s)$ Before and After the Time Delay}
\label{sec:psi_timedelay}
As we discussed in section~\ref{Populations Merging}, the large deviation function $\Psi(s)$ can be recovered from the slope in time of the logarithm of the average population. We also mentioned in section~\ref{Discreteness Effects at Initial Times}, an alternative we can consider to overcome the discreteness effects would be to eliminate the initial transient regime  where these effects are strong. As we will synthesize later, the improvement in the estimation of $\Psi(s)$ comes precisely from these two main contributions, the time delaying of populations and the discarding of the initial transient regime of the populations.

Let's call $\Psi(s)$ the analytical prediction for the large deviation function (given by equation~(\ref{eq:9})). $\Psi_{\num}(s)$ is obtained from the slope of the  logarithm of the average population (computed from merging several populations that have been generated using the cloning algorithm). $\Psi_{\tau}(s)$ is obtained through a time delay procedure over $\hat{N}$, as was described above. These two numerical estimations are in fact averages over $R$ realizations and over their last $\gamma$ values. The method how $\Psi_{\num}(s)$ and $\Psi_{\tau}(s)$ are computed is explained below.

\subsection{Numerical Estimators for $\Psi(s)$}

\begin{figure}[h!]
	\centering
        {\includegraphics [scale=0.7]{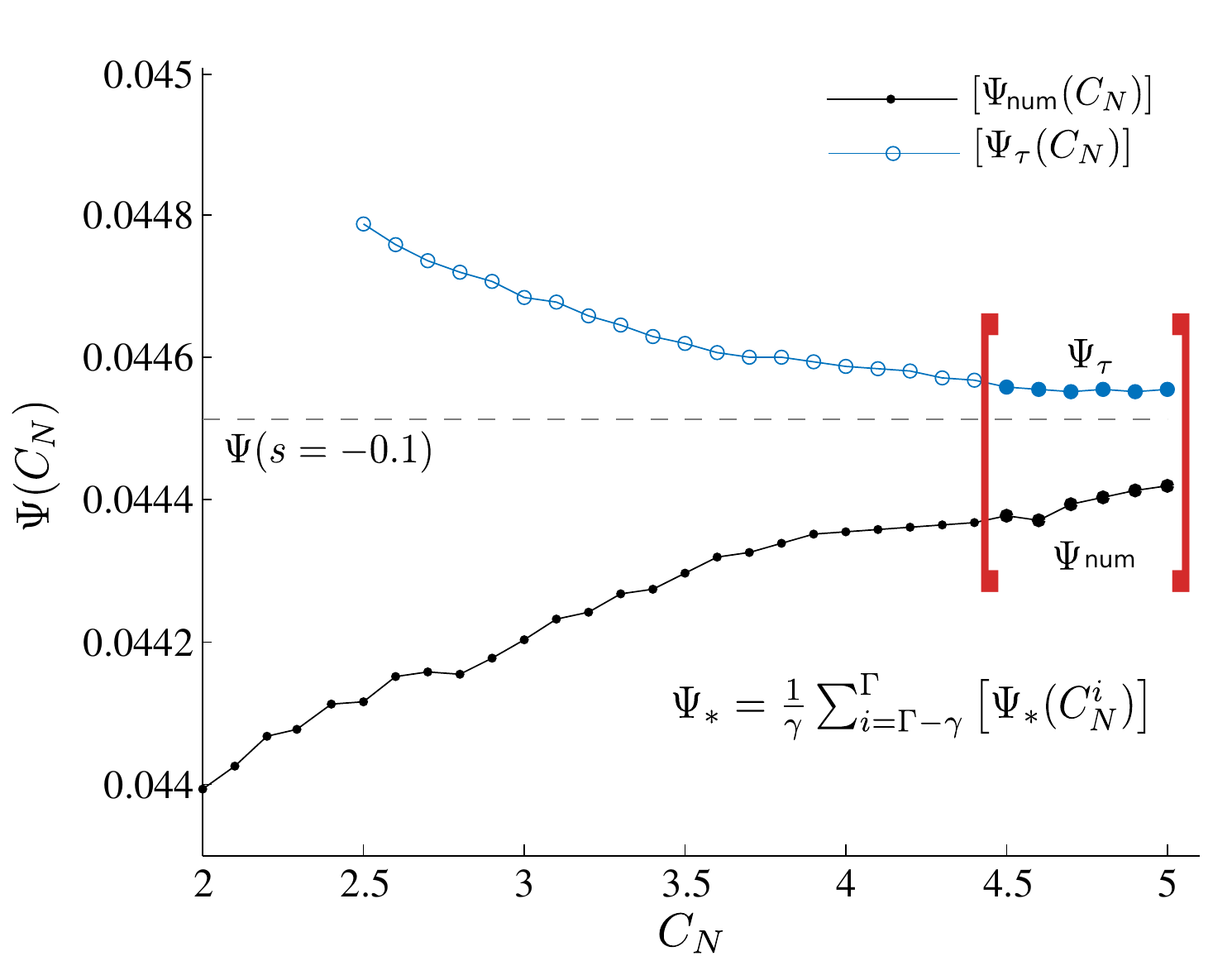}}	
	\centering        	
\caption{Numerical estimations of $\Psi(s=-0.1)$ as a function of the cut $C_{N}$ in (log) population. $\left[ \Psi_{\tau}(C_{N}) \right]$ is shown in blue and $\left[ \Psi_{\num}(C_{N}) \right]$ in black for $R = 40$. The numerical estimations $\Psi_{\num}$ and $\Psi_{\tau}$ are computed from an average of $\left[ \Psi_{*}(C_{N}^{i}) \right]$  over its last $\gamma = 6$ values. The subscript $``*"$ denotes ``$\num$'' or ``$\tau$''.}
\label{fig:psi1}
\end{figure}

Let's call $\Psi_{*}(C_{N})$ an estimation of $\Psi$ (by some method $(*)\in\{ \num, \tau \}$) as a function of the cut in log-population $C_{N}$. If we consider $C_{N}$ as $C_{N} = \{ C^{1}_{N},...,C^{\Gamma}_{N} \}$ a set of $\Gamma$ cuts, $\Psi_{*}(C_{N})$ is in fact $\Psi_{*}(C_{N}) = \{ \Psi_{*}(C^{1}_{N}),...,\Psi_{*}(C^{\Gamma}_{N}) \}$. If $\left[ \Psi_{*}(C^{i}_{N}) \right]$ is an average over $R$ realizations,

\begin{equation}
\left[ \Psi_{*}(C_{N}^{i}) \right] = \frac{1}{R} \sum_{r =1}^{R} \Psi_{*}^{r}(C_{N}^{i})
\end{equation}
our numerical estimation (for a given $s$) is then computed from an average of $\left[ \Psi_{*}(C_{N}^{i}) \right]$  over its last $\gamma$ values, \emph{i.e.},

\begin{equation}
\Psi_{*}(s)=\frac{1}{\gamma}
 \sum_{i=\Gamma-\gamma}^{\Gamma} \left[ \Psi_{*}(C_{N}^{i}) \right] =\frac{1}{\gamma R}
 \sum_{i=\Gamma-\gamma}^{\Gamma} \sum_{r =1}^{R} \Psi_{*}^{r}(C_{N}^{i})
\end{equation}
as is shown in figure~\ref{fig:psi1}. More details of the determination of these estimators are given in the subsection below.

\subsection{Comparison between``Bulk" and ``Fit" estimators of $\Psi(s)$}
\label{``Bulk" and ``Fit" Slopes}
\begin{figure}[h!]
        \centering
        \subfigure[``Bulk" slope] {\includegraphics [scale=0.51] {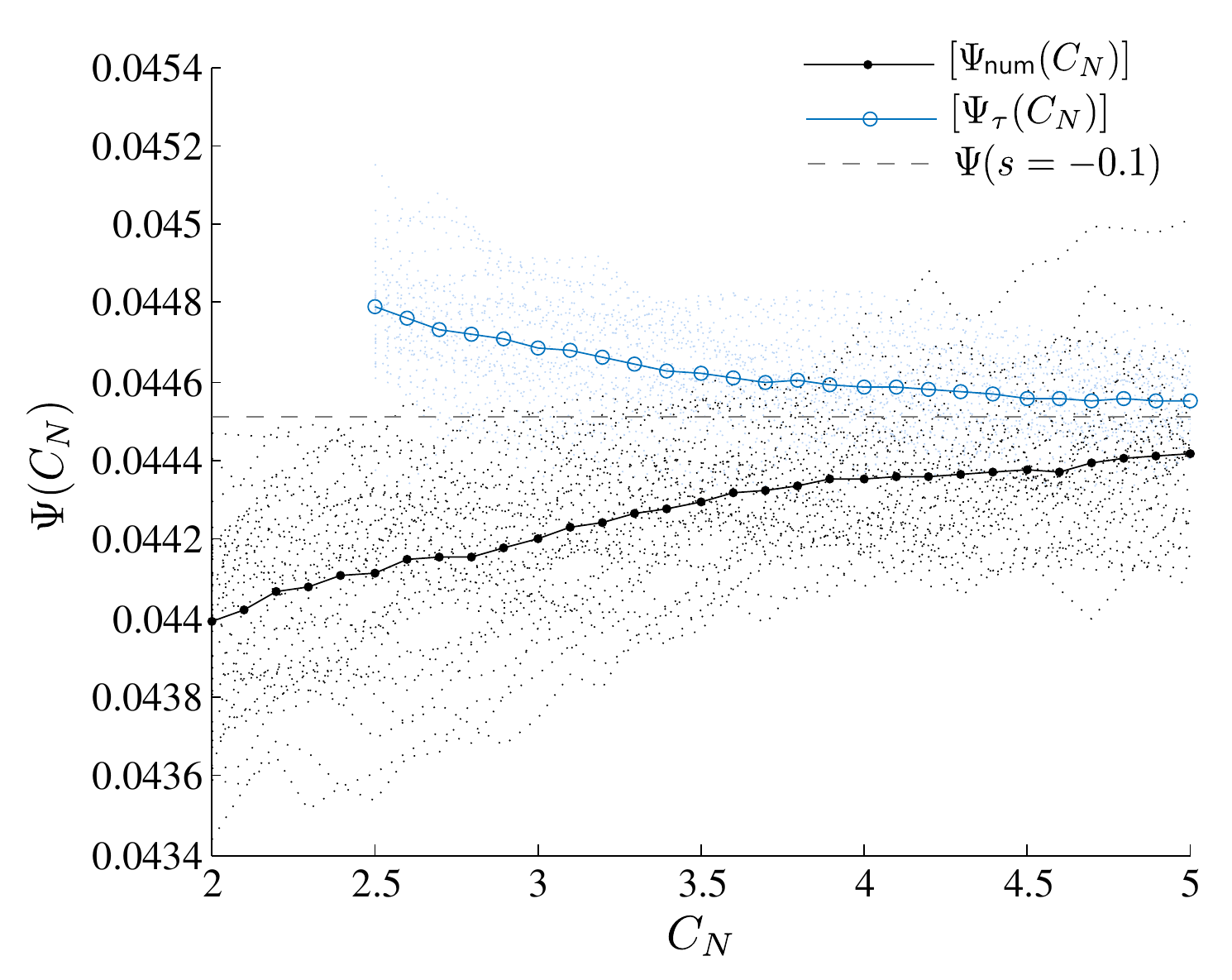}}
		\subfigure[``Fit" slope]{\includegraphics [scale=0.51] {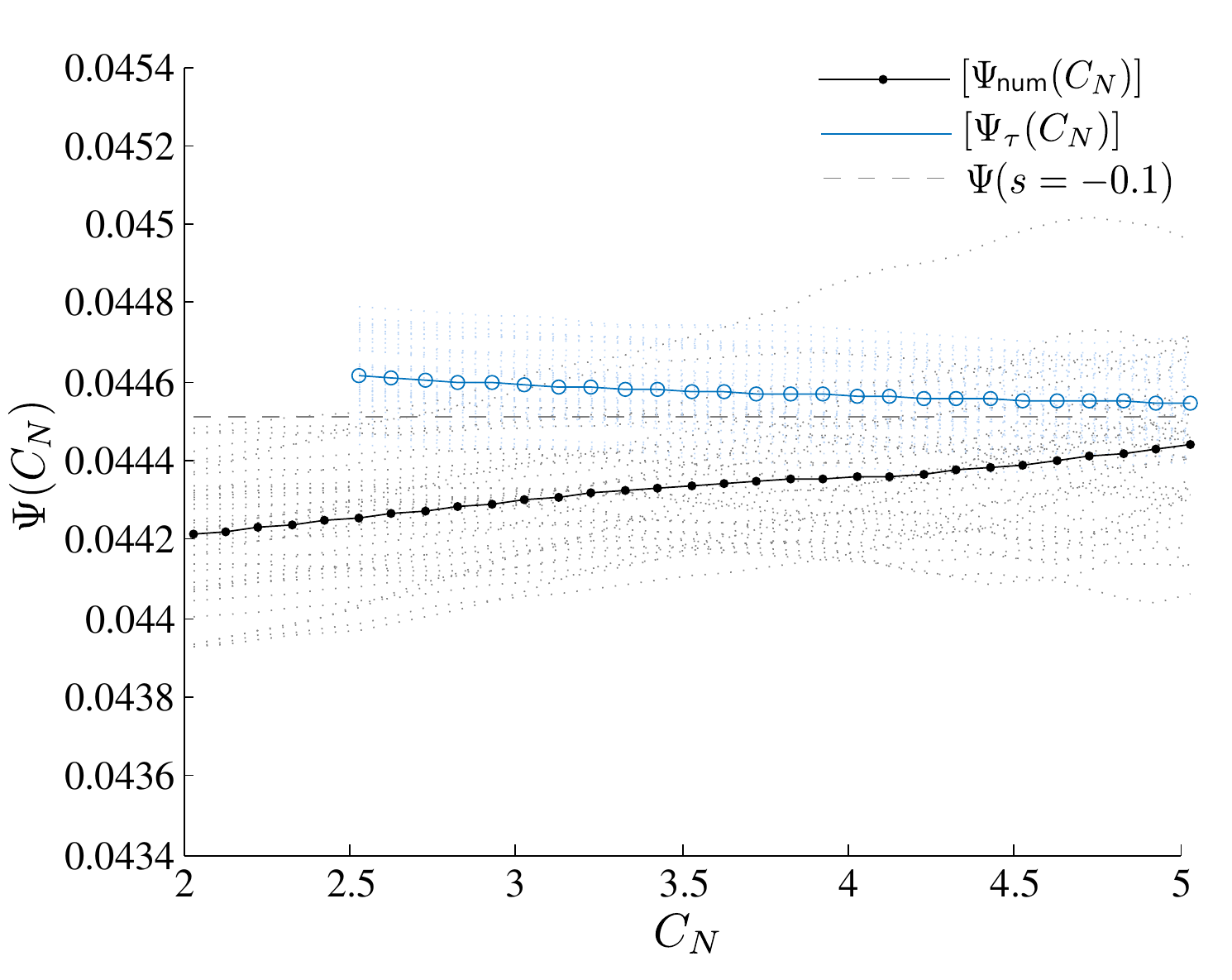}}\\ 
\centering        	
\caption{Average over $R = 40$ realizations of the numerical estimators $\Psi_{\num}(C_{N})$ and $\Psi_{\tau}(C_{N})$ as a function of the cut in log-population for $s=-0.1$. The estimation for $\Psi$ becomes better if we discard the initial instances where discreteness effects are strong.}
\label{fig:psi2}
\end{figure}

The estimators we defined in the last subsection can be obtained from the ``bulk" slope (figure~\ref{fig:psi2}(a)) given by equation~(\ref{eq:6})  and from the slope that comes from the affine fit of the average log-population by $\hat N(s,t) = mt + b$ (figure~\ref{fig:psi2}(b)). 
Figure~\ref{fig:psi2} shows the average over $R = 40$ realizations of the numerical estimators $\Psi_{\num}(C_{N})$ and $\Psi_{\tau}(C_{N})$ as a function of the cut in log-population for $s=-0.1$. As before, $\left[ \Psi_{\tau}(C_{N}) \right]$ is shown in blue and $\left[ \Psi_{\num}(C_{N}) \right]$ (without the ``time delay") is shown in black. As we already mentioned, the estimation for $\Psi$ becomes better if we discard the initial transient regime where the discreteness effects are strong. 

The black curves in figure~\ref{fig:psi2} represent the standard way of estimating $\Psi$ which come from the slope of the average (log) population, shown in dark green in figure~\ref{fig:delayedPOP}(a) for one realization. We can observe the effect of discarding the initial transient regime of these populations by cutting systematically this curve and computing $\Psi_{\num}(C_{N})$ from the growth rate $m$ computed on the interval $[C_{N},N_{\max}]$. Independently if $\Psi_{\num}(C_{N})$ is computed from the ``bulk" slope or by the ``fit" slope, for appropriate values of $C_{N}$, $\Psi_{\num}(C_{N})$ becomes closer to the theoretical value. Additionally to this result, we can add the ``time correction" or ``delay" we proposed in section~\ref{Time Delay Correction} and as can be seen in the blue curves in figure~\ref{fig:psi2}, the estimation $\Psi_{\tau}(C_{N})$ is even better and closer to the theoretical value  than  $\Psi_{\num}(C_{N})$ for all $C_{N}$.

Once we have proved that the estimation of $\Psi$ becomes better when we discard the initial times where the discreteness effects are strong and when we perform a ``time delay" over our populations in order to give more weight to the final instances of our populations, the question that remains is related to what we should consider as $\Psi_{\num}(s)$ and $\Psi_{\tau}(s)$. As we showed in figure~\ref{fig:psi1}, $\Psi_{\num}(s = -0.1)$ and $\Psi_{\tau}(s = -0.1)$ are computed from an average over the last $\gamma$ values of $\left[ \Psi_{\tau}(C_{N}) \right]$ and $\left[ \Psi_{\num}(C_{N}) \right]$. Below, we repeat this procedure and compute these estimators for several values of $s$, $s \in \left[ -0.3,-0.05 \right]$. The improvement in the determination of the ldf is measured through the relative distance of the numerical estimations with respect to the theoretical values and their errors.

\subsection{Relative Distance and Estimator Error}
\label{Relative Distance and Estimator Error}
\begin{figure}[h!]
        \centering
        \subfigure[``Bulk" slope] {\includegraphics [scale=0.51] {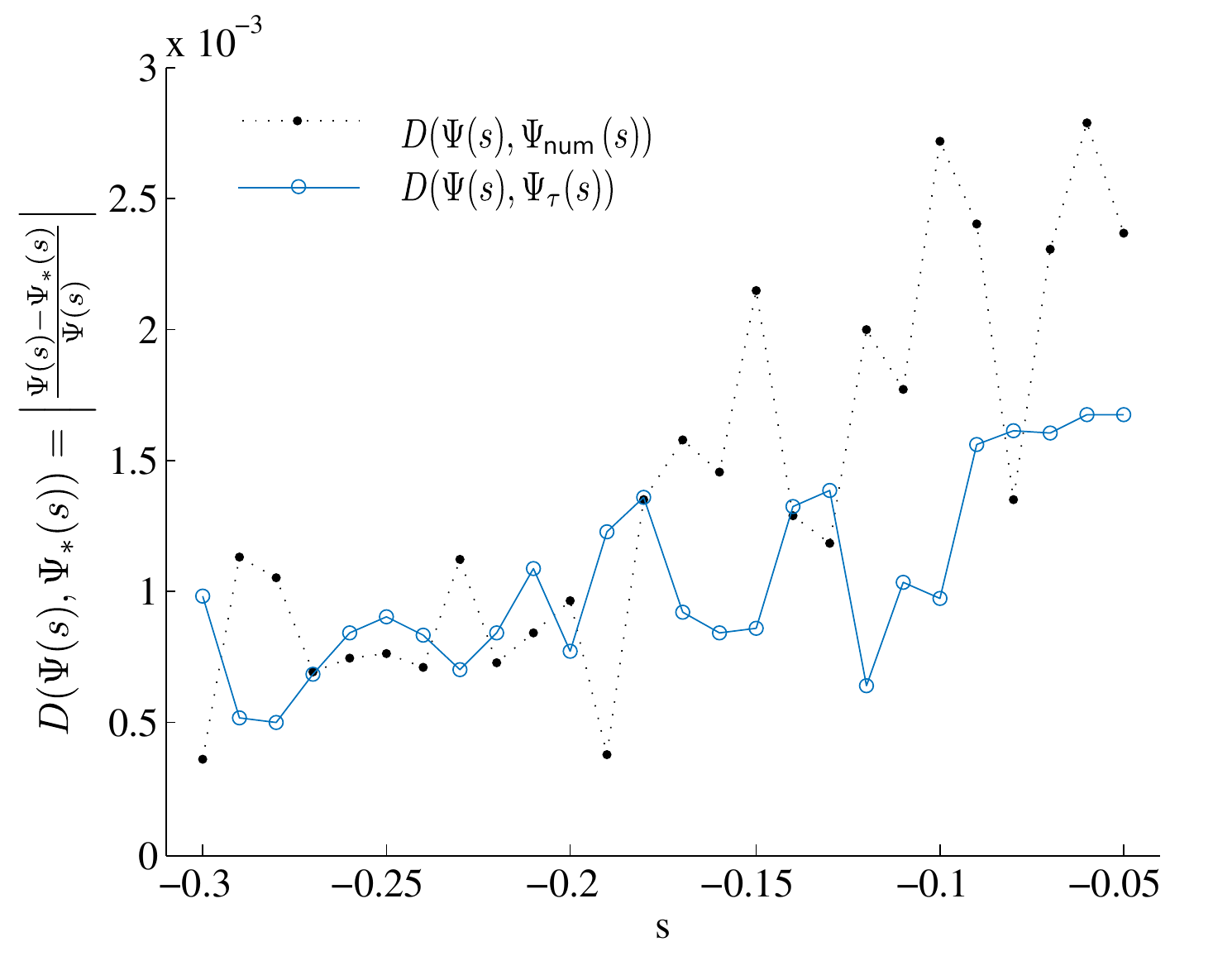}}
		\subfigure[``Fit" slope]{\includegraphics [scale=0.51] {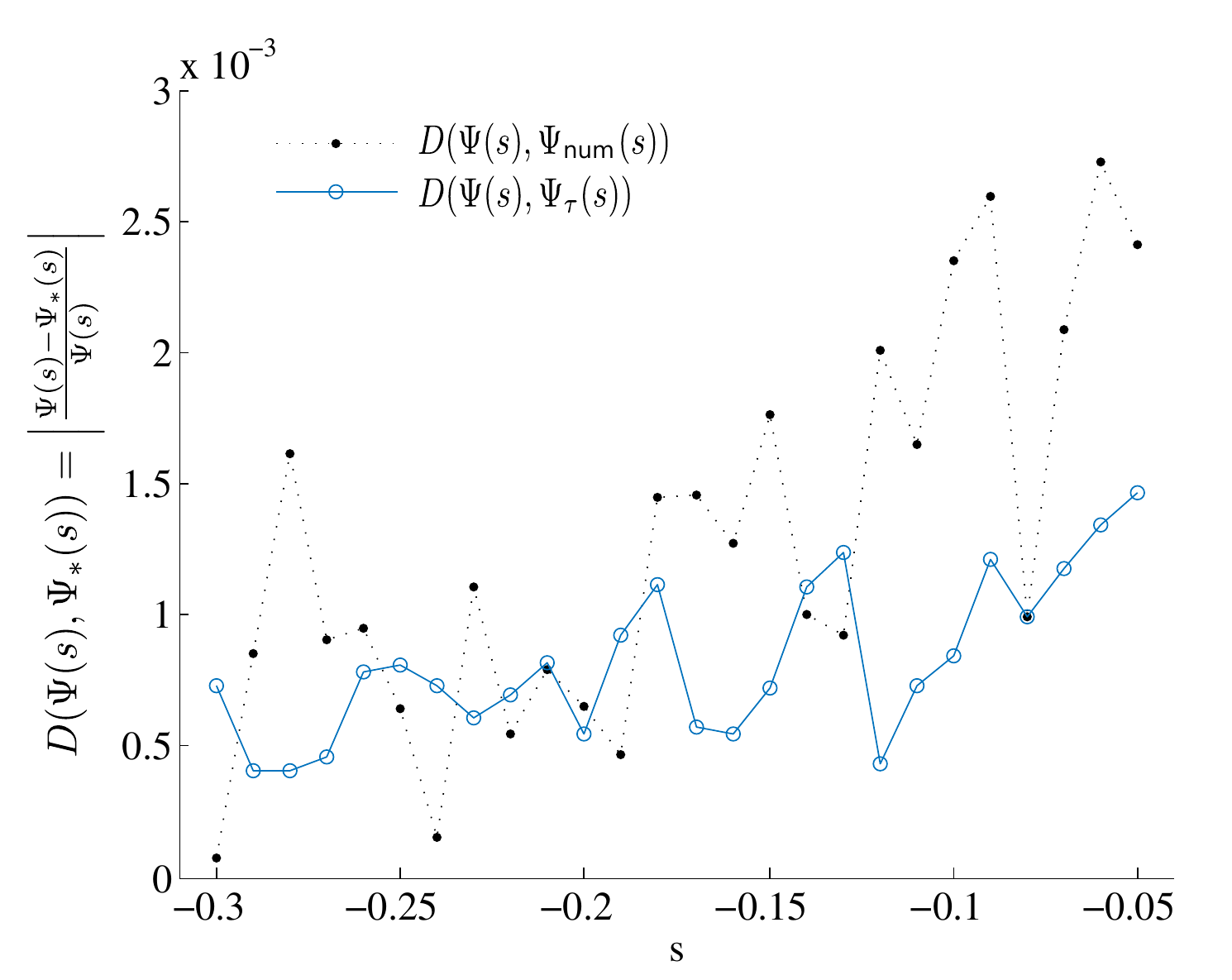}}\\ 
\centering        	
\caption{Relative distance $D(\Psi(s),\Psi_{*}(s))$ between the estimator $\Psi_{*}(s)$ and the theoretical value $\Psi(s)$. The deviation from the theoretical value is larger for values of $s$ close to $0$, but is smaller after the ``time delay correction'' for almost every value of $s$}.
\label{fig:relativeD}
\end{figure}

The relative distance
\begin{equation}
D(\Psi(s),\Psi_{*}(s)) = \left \vert \frac{ \Psi(s) - \Psi_{*}(s)}{\Psi(s)}  \right \vert
\end{equation}
between the estimator $\Psi_{*}(s)$ and the theoretical value $\Psi(s)$ is shown in figure~\ref{fig:relativeD}.

These distances were also computed from the ``bulk'' (a) and the ``fit'' slope (b) and with (blue) and without time delay (black). As we can observe, the deviation from the theoretical value is larger for values of $s$ close to $0$, but is smaller after the ``time correction'' for almost every value of $s$.

\begin{figure}[h!]
        \centering
        \subfigure[``Bulk" slope] {\includegraphics [scale=0.51] {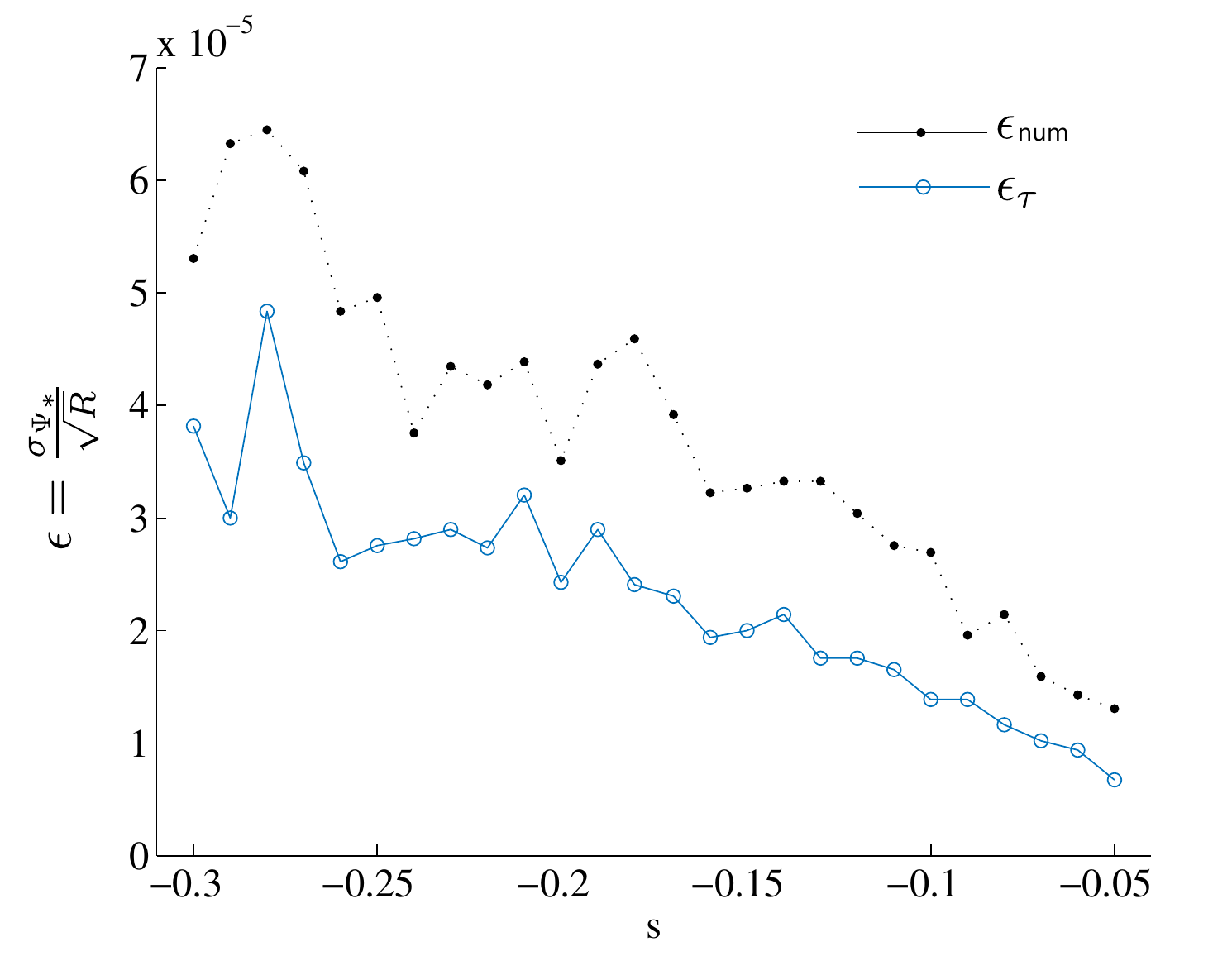}}
		\subfigure[``Fit" slope]{\includegraphics [scale=0.51] {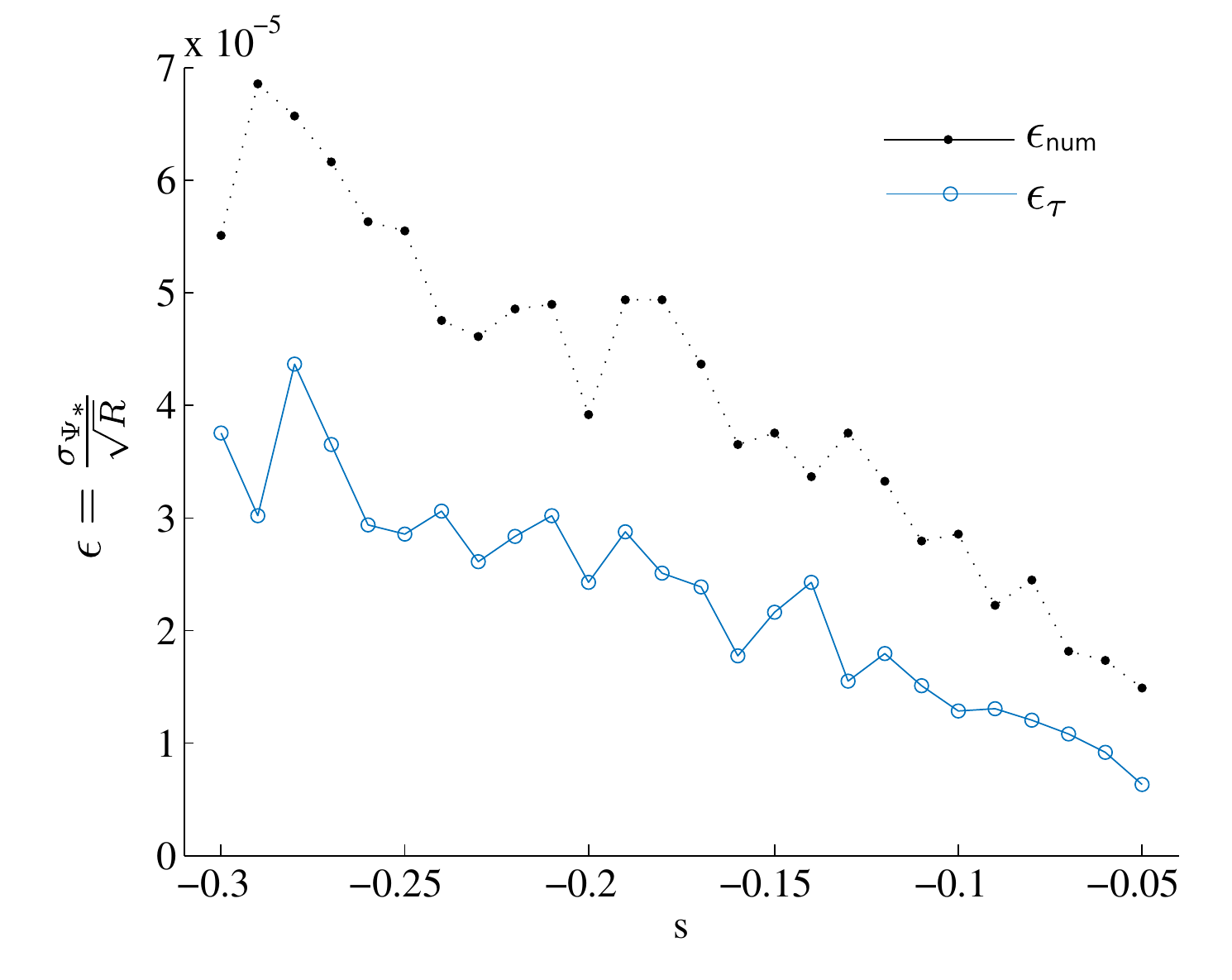}}\\ 
\centering        	
\caption{Estimator error for $\Psi(s)$, $\epsilon_{num}$ (black) and $\epsilon_{\tau}$ (blue). The estimator error decreases as $s$ approaches to $0$ (for both slopes) and it is always smaller for $\Psi_{\tau}(s)$ for any value of $s$.}.
\label{fig:error}
\end{figure}

In figure~\ref{fig:error} we present the estimator error for $\Psi(s)$ defined as
\begin{equation}
\epsilon = \frac{\sigma_{\Psi_{*}}}{\sqrt{R}} 
\end{equation}
where $R$ is the number of realizations and $\sigma_{\Psi_{*}}$ is the standard deviation of $\Psi_{*}(s)$. Similarly as in previous results, the estimator error decreases as $s$ approaches to $0$ (for both slopes) and it is always smaller for $\Psi_{\tau}(s)$ for any value of $s$.

\section{Time Delay Properties}
\label{sec:time-delay-prop}
In this section we analyse properties of the distribution of time delays  $\Delta\tau(s) = \{ \Delta\tau_{1}(s),...,\Delta\tau_{J}(s) \}$. This distribution has been centered with respect to its mean. 

\begin{figure}[h!]
        \centering
        \subfigure[Time Delay Variance] {\includegraphics [scale=0.51] {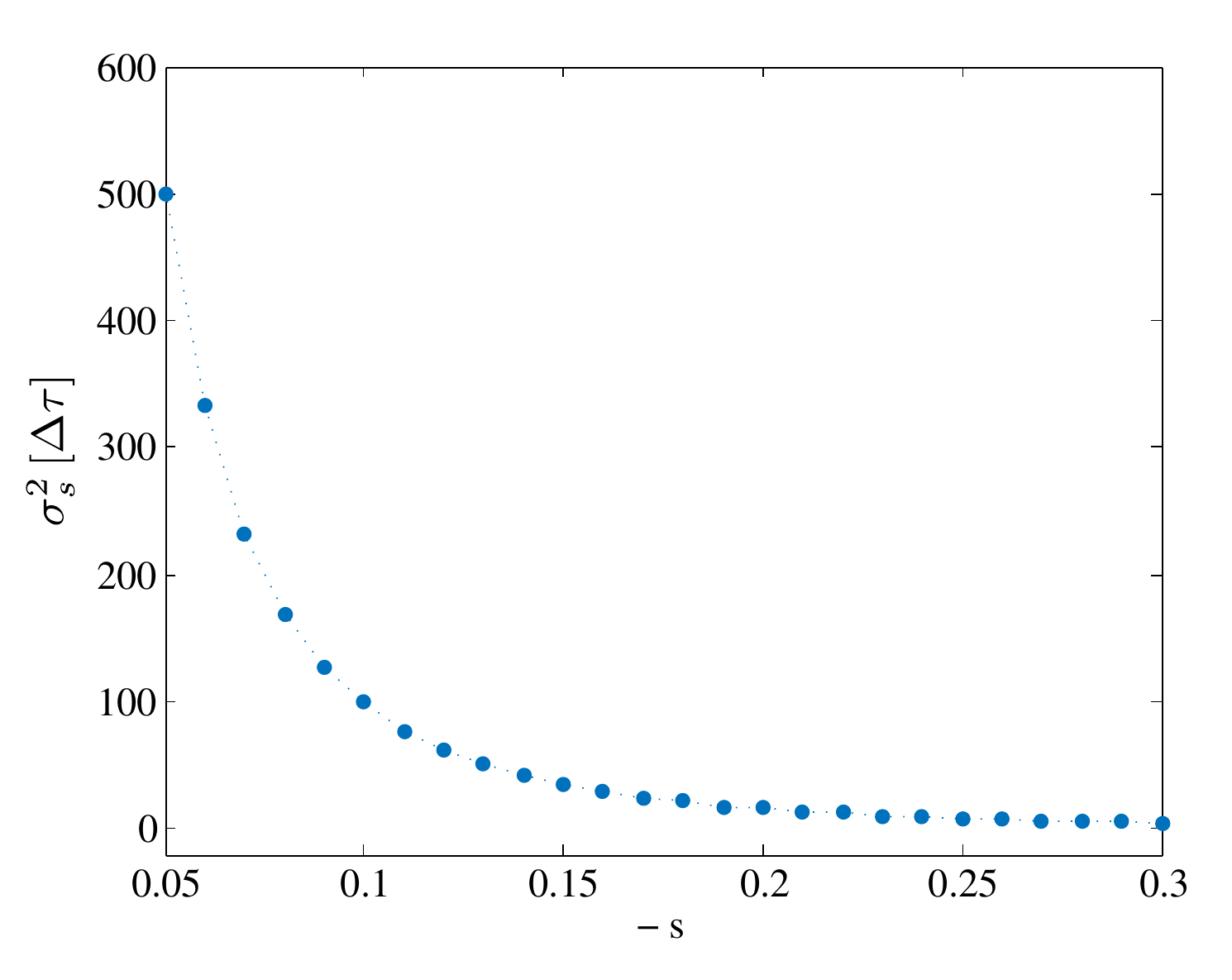}}
		\subfigure[Time Delay Variance in log-log scale]{\includegraphics [scale=0.51] {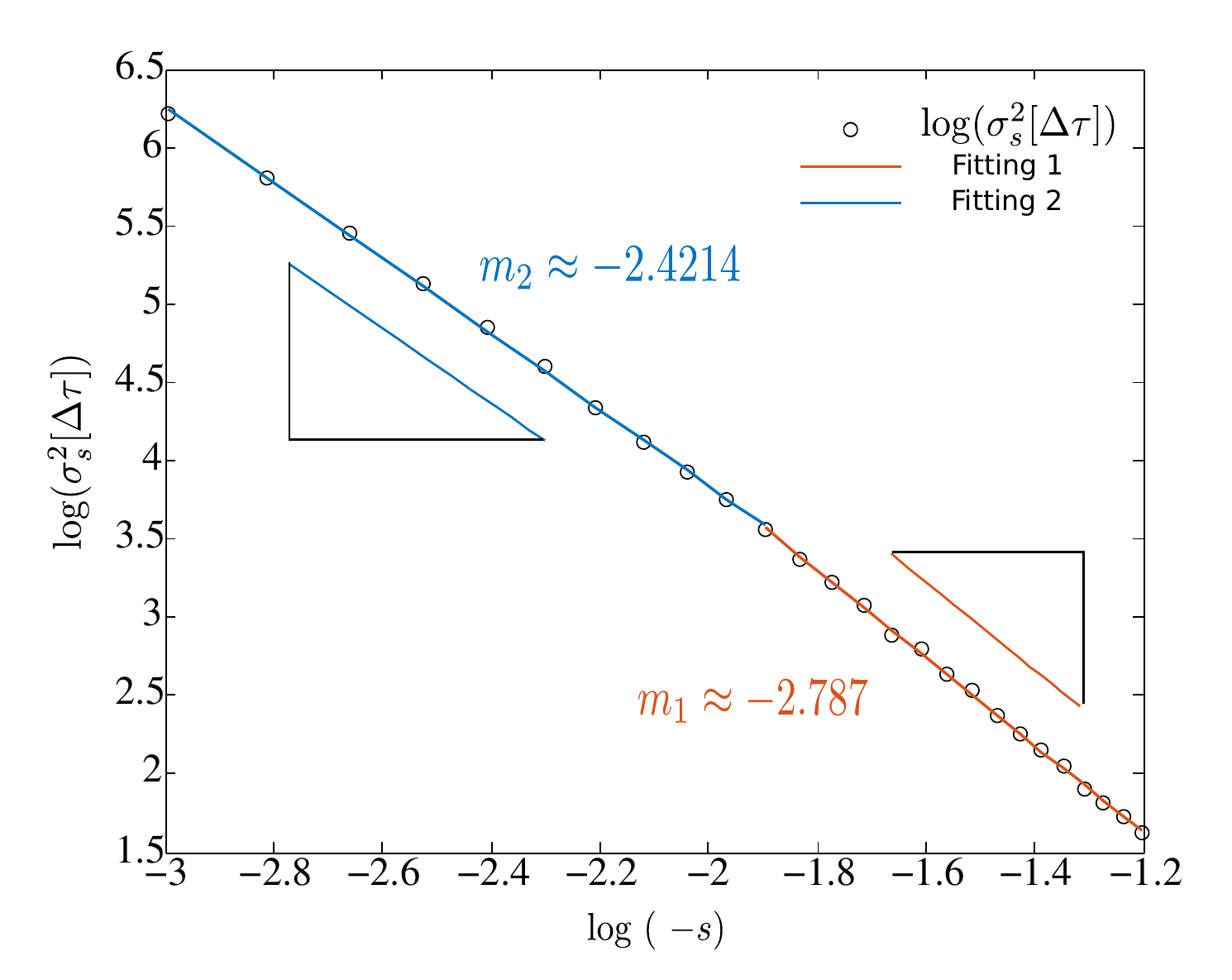}}\\ 
\centering        	
\caption{(a) Time delay variance $\sigma^{2}_{s} \left[ \Delta \tau \right]$. The dispersion of the time delays is large for values of $s$ close to $0$ and decreases rapidly as $-s$ increases. (b) Time delay variance regimes, one characterized with $m_{1}\approx -2.877$ ($s \in \left[ -0.15,-0.3 \right]$) and the other with $m_{2}\approx -2.4214$ ($s \in \left[ -0.05,-0.15 \right]$).}
\label{fig:timedelay}
\end{figure}

In figure~\ref{fig:timedelay}(a), we show the variance $\sigma^{2}_{s} \left[ \Delta \tau \right]$ of the time delay distribution $\Delta\tau(s)$. As we can see the dispersion of time delays is large for values of $s$ close to $0$ and decreases quickly as $-s$ increases. This is understood by observing that the typical growth rate $\left[ r_{s}(C) - r(C) \right] $ of the cloning algorithm goes to zero as $s \rightarrow 0 $ inducing a longer transient regime between the small and large population regimes. When we plot the variance in log-log scale, as in figure~\ref{fig:timedelay}(b), we can observe two linear regimes, one characterized by an exponent $m_{1}\approx -2.877$ ($s \in \left[ -0.15,-0.3 \right]$) and the other by $m_{2}\approx -2.4214$ ($s \in \left[ -0.05,-0.15 \right]$). They correspond to power-law behaviours in time of the variance of the delays, which remain to be understood.

\begin{figure}[h!]
	\centering
        \subfigure[Time Delay Distribution] {\includegraphics [scale=0.51] {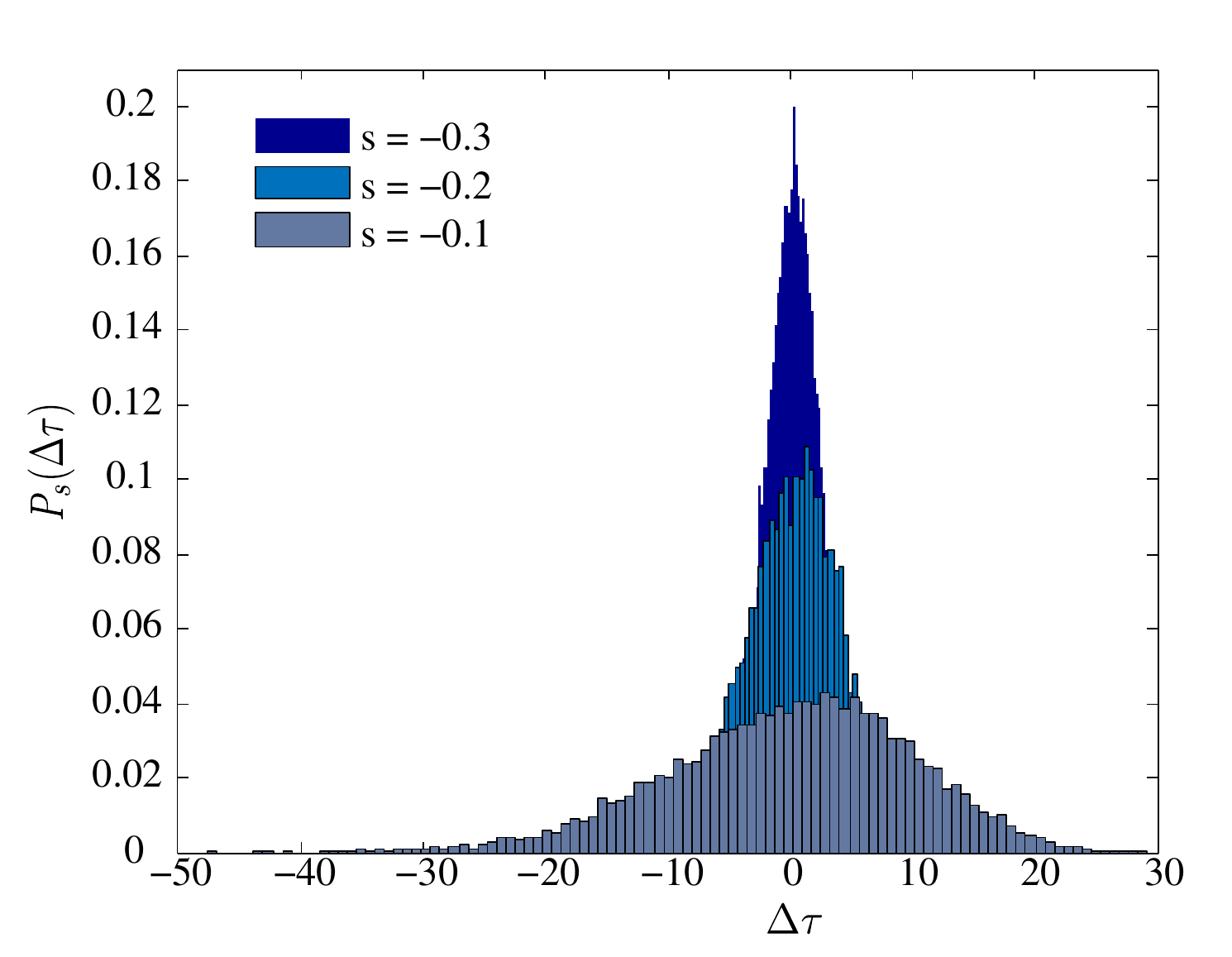}}
		\subfigure[ Rescaled Distribution $\hat{P} \left(\frac{\Delta \tau }{\sigma_{s} \left[\Delta\tau \right]}\right)$]{\includegraphics [scale=0.51] {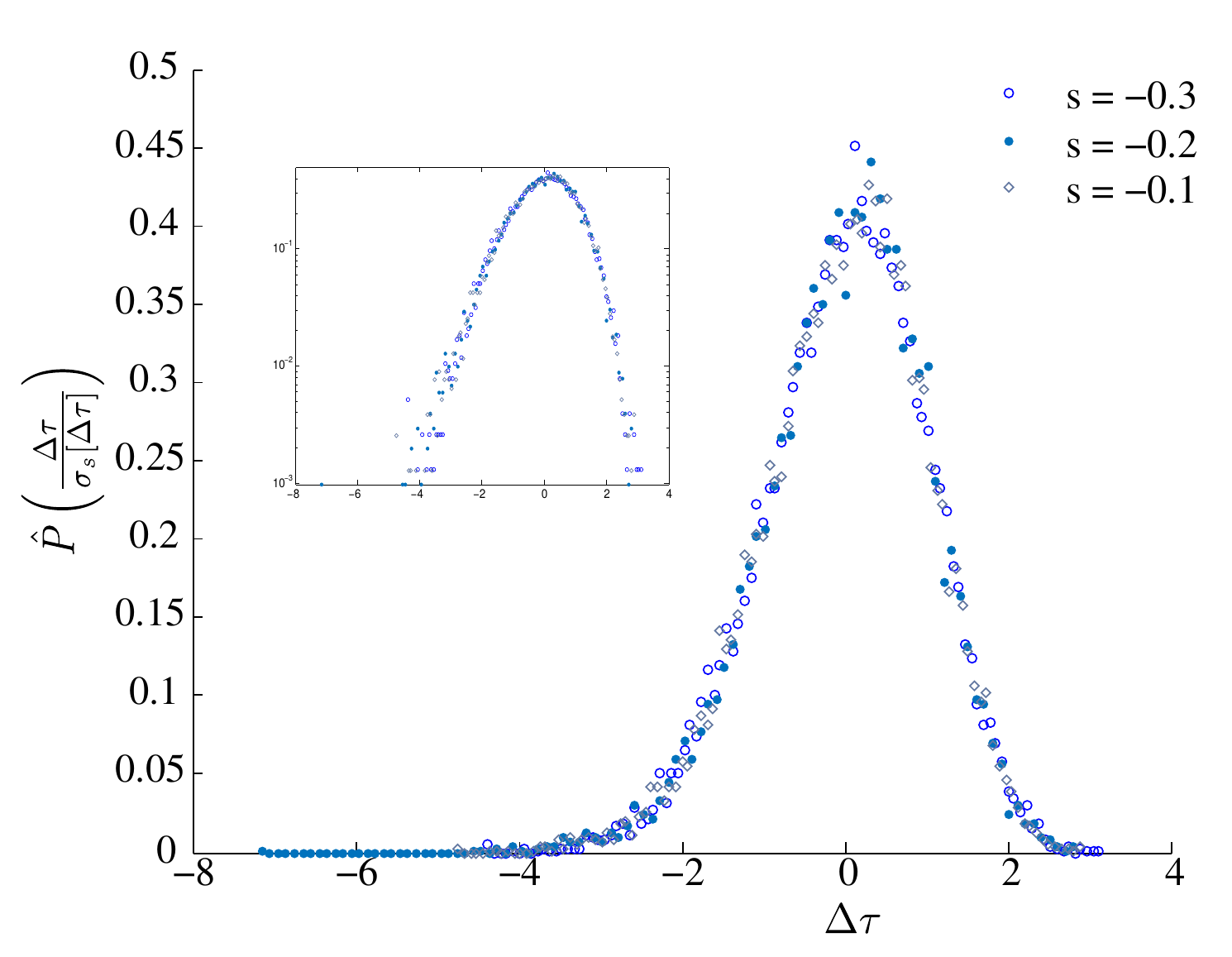}}\\ 
	\centering        	
\caption{(a) Distribution of time delays for different values of $s$. The dispersion of time delays is wider for values of $s$ closer to zero. (b) Rescaled distribution of time delays $\hat{P} \left(\frac{\Delta \tau }{\sigma_{s} \left[\Delta\tau \right]}\right)$. The distribution of time delays depends only on their $\sigma_{s}$ as $P_{s}(\Delta\tau) = \sigma_{s} \left[\Delta\tau \right] \hat{P} \left(\frac{\Delta \tau }{\sigma_{s} \left[\Delta\tau \right]}\right)$. }
\label{fig:histogram}
\end{figure}

This dependence of the dispersion of time delays with $s$ can be better seen in the distribution of time delays $P_{s}(\Delta\tau)$ shown in figure~\ref{fig:histogram} for various values of $s$. This distribution is wider for values of $s$ closer to zero (figure~\ref{fig:histogram}(a)). However if we rescale the distributions of time delays by their respective $\sigma_{s}$, as shown in figure~\ref{fig:histogram}(b), the distributions become independent of $s$ as $P_{s}(\Delta\tau) = \sigma_{s} \left[\Delta\tau \right] \hat{P} \left(\frac{\Delta \tau }{\sigma_{s} \left[\Delta\tau \right]}\right)$. This provides a strong numerical evidence supporting the existence of a universal distribution $\hat P$.

\section{Discussion}
\label{sec:discussion}
In this paper we analysed the discreteness effects at initial times in population dynamics. During the initial transient regime of the evolution of populations, there is a wide distribution of times at which the first series of jumps occurs. This means that fluctuations at initial times produce that some populations remain in their initial states for much longer than others, producing a gap in their individual evolution. This induces a relative shift that lasts forever. These effects play an important role specially for the determination of the large deviation function which may be obtained from the growth rate of the average log-population (section~\ref{The Cloning Algorithm and the Large Deviation Function}).

However, in section~\ref{Discreteness Effects at Initial Times} we saw how by restricting the evolution of our populations up to a maximum time $T_{\max}$ (or population $N_{\max}$) which is not ``large enough'', the average population (and $\Psi(s)$) is strongly affected by the behaviour of $\mathcal{N}$ at initial times. We proposed as an alternative to overcome the influence of initial discreteness effects to get rid of the regions of the populations where  these effects are present. In other words, to cut the initial transient regime of the populations. In that case, we saw that the average of populations is restricted to the interval $[\max \mathcal{T_{C}}, \min \mathcal{T_{F}}]$ which can be in fact very small and this can induce a poor estimation of $\Psi(s)$ (Figure~\ref{fig:merge}(b)).

Complementary to this, we found a way of emphasizing the effects of the exponential growth regime in the determination of $\Psi(s)$ by using the fact that log-populations after a long enough time become parallel (figure~\ref{fig:distance1}(a)) and that once the populations have overcome the discreteness effects, the distance between them becomes constant (figure~\ref{fig:distance1}(b)) and the discreteness effects are not strong anymore (section~\ref{Parallel Behaviour in Log-Populations}). We argue in section~\ref{Time Delay Correction} that this initial discreteness effects or initial ``lag'' between populations could be compensated by performing over the populations a time translation (equation (\ref{eq:13})). This time delay procedure is chosen so as to overlap the population evolutions in their large-time regime (figure~\ref{fig:delayedPOP}(b)). The improvement in the estimation of $\Psi$ comes precisely from these two main contributions, the time delaying of populations and the discarding of the initial transient regime of the populations. 

We showed how the the numerical estimations for the ldf are improved as the initial instances of the populations are discarded (independently of the method used to compute the growth rate of the average population, see figure~\ref{fig:psi2}). Also, it is was shown that if additionally, we perform the time delay procedure, the estimation of $\Psi$ is improved even more and closer to the theoretical value (section~\ref{``Bulk" and ``Fit" Slopes}). This result was confirmed later in section~\ref{Relative Distance and Estimator Error} by computing the relative distance of the numerical estimators with respect to the theoretical value and their errors. As we observed (figure~\ref{fig:relativeD}), the deviation from the theoretical value is higher for values of $s$ close to $0$, but is smaller after the ``time correction'' for almost every value of $s$. Similarly for the error estimator (figure~\ref{fig:error}).

Our numerical study was performed on a simple system, and we hope it can be extended to more complex phenomena. However, there remain open questions even for the death-and-birth system we have studied. The duration of the initial discrete-population regime could be understood from an analytical study of the population dynamics itself. Our numerical results also support a power-law behaviour in time of the variance of the delays. Furthermore, it appeared that the distribution of the delays takes a universal form, after rescaling the variance to one. Those observations open questions for future studies.

\section*{Acknowledgements}
Esteban Guevara thanks Khashayar Pakdaman for his support and discussions. This project was partially funded by the PEPS LABS and the LAABS Inphyniti CNRS projects. Special thanks to the Ecuadorian Government and the Secretar\'ia Nacional de Educaci\'on Superior, Ciencia, Tecnolog\'ia e Innovaci\'on, SENESCYT.

\end{document}